\documentclass{article}
\usepackage[utf8]{inputenc}
\usepackage[margin=1in]{geometry}
\usepackage{macros}
\usepackage{authblk}

\title{Lower Bounds for Set-Multilinear Branching Programs}
\author[1]{Prerona Chatterjee\footnote{Research supported by the Azrieli International Postdoctoral Fellowship, the Israel Science Foundation (grant number 514/20) and the Len Blavatnik and the Blavatnik Family foundation. Email: prerona.ch@gmail.com}}
\author[2]{Deepanshu Kush\footnote{Email: deepkush@cs.toronto.edu}}
\author[2,3]{Shubhangi Saraf\footnote{Research partially supported by a Sloan research fellowship and an NSERC Discovery Grant. Email: shubhangi.saraf@utoronto.ca}}
\author[1]{Amir Shpilka\footnote{Research leading to these results has received funding from the  Israel Science Foundation (grant number 514/20) and from the Len Blavatnik and the Blavatnik Family foundation. Email: shpilka@tauex.tau.ac.il}}

\affil[1]{Blavatnik School of Computer Science, Tel-Aviv University}
\affil[2]{Department of Computer Science, University of Toronto}
\affil[3]{Department of Mathematics, University of Toronto}

\date{\today}

\begin{document}
\maketitle

\begin{abstract}
   
In this paper, we prove super-polynomial lower bounds for the model of \emph{sum of ordered set-multilinear algebraic branching programs}, each with a possibly different ordering ($\sbp$). 
Specifically, we give an explicit $nd$-variate polynomial of degree $d$ such that any $\sbp$ computing it must have size $n^{\omega(1)}$ for $d$ as low as $\omega(\log n)$. Notably, this constitutes the first such lower bound in the low degree regime. Moreover, for $d = \poly(n)$, we demonstrate an exponential lower bound.
This result generalizes the seminal work of Nisan (STOC, 1991), which proved an exponential lower bound for a single ordered set-multilinear ABP.

The significance of our lower bounds is underscored by the recent work of Bhargav, Dwivedi, and Saxena (to appear in TAMC, 2024), which showed that super-polynomial lower bounds against a sum of ordered set-multilinear branching programs -- for a polynomial of sufficiently low degree -- would imply super-polynomial lower bounds against general ABPs, thereby resolving Valiant's longstanding conjecture that the permanent polynomial can not be computed efficiently by ABPs. 
More precisely, their work shows that if one could obtain such lower bounds when the degree is bounded by $O(\log n/ \log \log n)$, then it would imply super-polynomial lower bounds against general ABPs. 

Our results strengthen the works of Arvind \& Raja (Chic. J. Theor. Comput. Sci., 2016) and Bhargav, Dwivedi \& Saxena (to appear in TAMC, 2024), as well as the works of Ramya \& Rao (Theor. Comput. Sci., 2020) and Ghoshal \& Rao (International Computer Science Symposium in Russia, 2021), each of which established lower bounds for related or restricted versions of this model.
They also strongly answer a question from the former two, which asked to prove super-polynomial lower bounds for general $\sbp$.

\end{abstract}

\newpage
\section{Introduction}
\subsection{Background on Algberaic Complexity}

In his seminal work (\cite{Val79}) in 1979, Valiant proposed an algebraic framework to study the computational complexity of computing \emph{polynomials}. \emph{Algebraic Complexity Theory} is this study of the complexity of computational problems which can be described as computing a multivariate polynomial $P(x_1,\ldots,x_N)$ over some elements $x_1,\ldots,x_N$ lying in a fixed field $\F$. Several fundamental computational tasks such as computing the determinant, permanent, matrix product, etc., can be represented using this framework. The natural computational models that we investigate in this setting are models such as \emph{algebraic circuits}, \emph{algebraic branching programs}, and \emph{algebraic formulas}. 

An \emph{algebraic circuit} over a field $\F$ for a multivariate polynomial $P(x_1,\ldots,x_N)$ is a directed acyclic graph (DAG)
whose internal vertices (called gates) are labeled as either $+$ (sum) or $\times$ (product), and leaves
(vertices of in-degree zero) are labeled by the variables $x_i$ or constants from $\F$. 
A special output gate (the root of the DAG) represents the polynomial
$P$. 
If the DAG happens to be a tree, such a resulting circuit is called an \emph{algebraic formula}. 
The size of a circuit or formula is the number of nodes in the DAG. We also consider the product-depth of the circuit, which is the maximum number of product gates on a root-to-leaf path. 
The class $\VP$ (respectively, $\VF$) is then defined to be the collection of all polynomials having at most polynomially large degree which can be computed by polynomial-sized circuits (respectively, formulas).

The class $\VP$ is synonymous to what we understand as efficiently computable polynomials. The class $\VNP$, whose definition is similar to the boolean class $\NP$, is in some sense a notion of what we deem as \emph{explicit}. Much like the problem of proving circuit size lower bounds for {explicit} boolean functions, the problem of proving them for explicit \emph{polynomials} (i.e., showing $\VP\neq \VNP$) has also remained elusive for many decades. However, because the latter only deals with {formal} \emph{symbolic} computation as opposed to modelling \emph{semantic} truth-table constraints, it is widely believed to be easier to resolve than its boolean counterpart. In fact, it is even known to be a \emph{pre-requisite} to the $\mathsf{P}\neq \NP$ conjecture in the non-uniform setting (\cite{burg}).

An \emph{algebraic branching program} (ABP) is a layered DAG with two special nodes in
it: a start-node and an end-node. All edges of the ABP go from layer $\ell -1$ to layer $\ell$ for some $\ell$ (say start-node is the unique node in layer $0$ and end-node is the unique node in the last layer) and are labeled by a linear polynomial. Every directed path $\gamma$ from start-node to end-node computes the monomial $P_\gamma$,
which is the product of all labels on the path $\gamma$. The ABP computes the polynomial $P = \sum_{\gamma} P_\gamma$,
where the sum is over all paths $\gamma$ from start-node to end-node. Its size is simply the number of nodes in the DAG, its \emph{depth} is the length of the longest path from the start-node to the end-node, and \emph{width} is the maximum number of nodes in any layer. The class $\VBP$ is then defined to be the collection of all polynomials (with polynomially-bounded degree) which can be computed by polynomial-sized branching programs. ABPs are known to be of intermediate complexity between formulas and circuits; in other words, we know the inclusions $\VF\subseteq\VBP \subseteq \VP\sse \VNP$.

It is conjectured that all of these inclusions are strict, and resolving any of these conjectures would represent a dramatic advancement in algebraic complexity theory, and even more broadly, in circuit complexity overall. Valiant's original hypothesis in \cite{Val79} pertains to showing a super-polynomial separation between the complexity of computing the determinant and the permanent polynomials.
This is known to be equivalent to the $\VBP \neq \VNP$ conjecture, i.e., showing super-polynomial size lower bounds against ABPs computing \emph{explicit} polynomials. 
At present, the best known lower bound against ABPs is only quadratic (\cite{ ChatterjeeKSV22}), and it appears as though we are quite distant from addressing this conjecture. On the other hand, as we now elaborate, while not directly improving upon this quadratic bound, this paper makes significant progress towards a different line of attack aimed at resolving Valiant's conjecture.

\subsection{Set-Multilinearity: A Key Syntactic Restriction}\label{subsec:sml}
One key advantage that algebraic models offer over their boolean counterparts is that of \emph{syntactic} restrictions. 
A recurring theme in algebraic complexity theory is to first efficiently convert general models of computation (such as circuits or formulas) to special kinds of syntactically-restricted models, show strong lower bounds against these restricted models, and then recover non-trivial lower bounds against the original general models owing to the efficiency of this conversion. This phenomenon is termed \emph{hardness escalation}. 
In this subsection, we describe one crucial example of a syntactic restriction in detail, that of \emph{set-multilinearity}.

A polynomial is said to be \emph{homogeneous}
if each monomial has the same total degree and \emph{multilinear} if every variable occurs
at most once in any monomial. Now, suppose that the underlying variable set is partitioned into $d$ sets
$X_1,\ldots, X_d$. Then the polynomial is said to be \emph{set-multilinear} with respect to this variable partition if each
monomial in $P$ has \emph{exactly} one variable from each set. Note that a set-multilinear polynomial is both multilinear and homogeneous, and has degree precisely $d$ if it is set-multilinear over $d$ sets. Next, we define different models
of computation corresponding to these variants of polynomials classes. An algebraic formula/branching program/circuit is set-multilinear with respect to a variable partition
$(X_1,\ldots, X_d)$ if each internal node in the formula/branching program/circuit computes a set-multilinear polynomial.\footnote{Of course, a non-root node need not be set-multilinear with respect to the \emph{entire} variable partition. Nevertheless, here we demand that it must be set-multilinear with respect to some \emph{subset} of the collection $\{X_1,\ldots,X_d\}$.} Multilinear and homogeneous formulas/branching programs/circuits are defined analogously.  

We now describe several important hardness escalation results, each reducing general models to corresponding \emph{set-multilinear} models.

\begin{paragraph}{Constant depth circuits.}
    The recent celebrated breakthrough work of Limaye, Srinivasan, and Tavenas (\cite{LST1}) establishes super-polynomial lower bounds for general algebraic circuits for \emph{all} constant-depths, a problem that was open for many decades. In order to show this, it is first shown that general low-depth algebraic formulas can be converted to set-multilinear algebraic formulas of low depth as well, and without much of a blow-up in size (as long as the degree is small). Subsequently, strong lower bounds are established for low-depth set-multilinear circuits (of small enough degree), which when combined with the first step yields the desired lower bound for general constant-depth circuits.
\end{paragraph}

\begin{paragraph}{Formulas.}
    Raz \cite{Raz-Tensor} showed that if an $N$-variate set-multilinear polynomial of degree $d$ has an algebraic formula of size $s$, then it also has a set-multilinear formula of size $\poly(s)\cdot (\log s)^{d}$. 
    In particular, for a set-multilinear polynomial $P$ of degree $d = O(\log N/ \log \log N)$, it follows that $P$ has a formula of size $\poly(N)$ if and only if $P$ has a set-multilinear formula of size $\poly(N)$. 
    Thus, having $N^{\omega_d(1)}$ set-multilinear formula size lower bounds for such a low degree would imply super-polynomial lower bounds for general formulas. A recent line of work by Kush and Saraf (\cite{KushS22,KushS23}) can be viewed as an attempt to
prove general formula lower bounds via this route. 

\end{paragraph}

\begin{paragraph}{Algebraic Branching Programs.}
    In fact, in the context of ABPs as well, the very recent work of Bhargav, Dwivedi, and Saxena (\cite{BDS23}) reduces the problem of showing lower bounds against general ABPs to proving lower bounds against sums of ordered set-multlinear ABPs (again, as long as the degree is small enough).
    Ordered set-multilinear ABPs are, in fact, historically well-studied models and extremely well-understood.
    However, despite their apparent simplicity, the work \cite{BDS23} implies that understanding their \emph{sums} -- a model that is far less studied -- is at the forefront of understanding Valiant’s conjecture.
    We state their result formally as \autoref{thm:hard-esc} in Section \ref{subsec:our-results}. 
    
    First however, as this is also the main model considered in this paper, we begin by formally defining \emph{ordered} set-multilinear ABPs and outlining their importance.     
\end{paragraph}

\begin{definition}[Ordered smABP]\label{def:smABP}
Given a variable partition $(X_1,\ldots, X_d)$, we say that a set-multilinear branching program of depth $d$ is said to be \emph{ordered} with respect to an 
ordering (or permutation) $\sigma\in S_d$ if for each $\ell\in[d]$, all edges of the ABP from layer $\ell-1$ to layer $\ell$ are labeled using a linear form over the variables in $X_{\sigma(\ell)}$. It is simply said to be \emph{ordered} if there \emph{exists} an ordering $\sigma$ such that it is ordered with respect to $\sigma$.
\end{definition}

At this point, it is essential to take note of the terminology in this context: in this paper, a general (or ``unordered'') set-multilinear branching program refers to an ABP for which each internal node computes a polynomial that is set-multilinear with respect to \emph{some} subset of the global partition, whereas an \emph{ordered} set-multilinear branching program is more specialized and has the property that any two nodes in the same layer compute polynomials that are set-multilinear with respect to the \emph{same} partition.

\begin{remark}
    This notion of ordered set-multilinear branching programs turns out to be equivalent to the more commonly used notions of (i) ``read-once oblivious algebraic branching programs (ROABPs)'', as well as (ii) ``non-commutative algebraic branching programs'' (see, for example, \cite{ForbesS13}).
    This relationship, especially with the former model, is described in more detail later in Section \ref{subsec:roabps}.
\end{remark}


\begin{definition}[$\sbp$]\label{def:sum}
Given a polynomial $P(X)$ that is set-multilinear with respect to the variable partition $X = (X_1,\ldots, X_d)$, we say that $\sum_{i=1}^t A_i$ is a $\sbp$ computing $P$ if indeed $\sum_{i=1}^t A_i(X) = P(X)$, and
each $A_i$ is an ordered set-multilinear branching program i.e., each $A_i$ is ordered with respect to some $\sigma_i\in S_d$. We call $t$ (i.e., the number of summands in a $\sbp$) its \emph{support size} and define its \emph{max-width} and \emph{total-width} to be the maximum over the width of each $A_i$ and the sum of the width of each $A_i$, respectively.    
\end{definition}

We have known \emph{exponential} width lower bounds against a \emph{single} ordered set-multilinear ABP since the foundational work of Nisan.
In \cite{Nisan91}, he showed that there are explicit polynomials (in fact, in $\VP$) which require any ordered set-multilinear ABP computing them to be of exponentially large width. 
Viewed differently, this work even shows that in the non-commutative setting, $\VBP \neq \VP$\footnote{We briefly explain the connection between ordered set-multilinear ABPs and non-commutative computation in Section \ref{subsec:roabps}.}. 
More crucially however, this work introduced a powerful technique -- a notion known as the \emph{partial derivative method} -- that has been instrumental in the bulk of the major advancements in algebraic complexity theory over the past three decades (such as \cite{NisanW97,Raz06, RazY09, Kayal12, KayalST16, KayalLSS17, /KumarS17, LST1, LST2}, see also \cite{ShpilkaY10,Saptarishi-survey}). 

Despite the considerable development of the partial derivative technique over the course of these works (and many more) for proving strong lower bounds against various algebraic models, relatively little is known about a \emph{general sum} of ordered set-multilinear ABPs -- a simple and direct generalization of the original model considered by Nisan. There is some progress in the literature towards this goal but which still requires additional structural restrictions on either the max-width or the support size or the size of each part in the variable partition.
The work \cite{ArvindR16} of Arvind and Raja shows that any $\sbp$ of support size $t$ computing the $n\times n$ permanent polynomial requires max-width (and therefore, total-width) at least $2^{\Omega(n/t)}$. 
Note that for this bound to be super-polynomial, the support size needs to be heavily restricted i.e., $t$ must be sub-linear. 
On the other hand, the work \cite{BDS23} also shows a super-polynomial lower bound in this context: it implies that no $\sbp$ of polynomially-bounded total-width can compute the iterated matrix multiplication (IMM) polynomial.
However, their work requires the additional assumption that the max-width of such an $\sbp$ is $n^{o(1)}$, that is \emph{sub-polynomial} in the number of variables. 

Apart from these, Ramya and Rao (\cite{RR20}) use the partial derivative method to show an exponential lower bound against the related model of sum of ROABPs in the multilinear setting, as well as some other structured multilinear ABPs.
Their lower bounds are for a multilinear polynomial that is computable by a small multilinear circuit.
Ghoshal and Rao (\cite{GR21}) partially extend their work by proving an exponential lower bound, for a polynomial that is computable even by a small multilinear ABP, against sums of ROABPs that have polynomially bounded width.
Notably, these results can be viewed as lower bounds against the $\sbp$ model where each variable set in the variable partition has size $2$ (that is, the total number of variables is $2d$).
This is because a multilinear polynomial and any multilinear model  computing it (such as circuit, formula, or branching program) can be converted, in a generic manner, to a set-multilinear polynomial and the corresponding set-multilinear model respectively, with each variable set having size $2$ (also see Section \ref{sec:RaghusResults} for a discussion). However, from the perspective of {hardness escalation of \cite{BDS23} that is described above} -- and which is indeed the focus of our work -- the setting of $d$ that is far more interesting is when it is allowed to be considerably smaller than $n$.
More precisely, the framework of \cite{BDS23} requires $d = O(\log n/\log \log n)$ (stated formally as Theorem \ref{thm:hard-esc} below).
A detailed discussion about the results in \cite{RR20}, \cite{GR21} and how they compare with our work can be found in Section \ref{sec:RaghusResults}.

\subsection{Our Results}\label{subsec:our-results}

Our main result is in this paper is a super-polynomial lower bound against an {unrestricted} sum of ordered set-multilinear branching programs, for a hard polynomial with ``small'' degree. We first state this result formally below, and then subsequently explain the connection with the hardness escalation result of \cite{BDS23} that is alluded to in the previous subsections.

\begin{theorem}[``Low''-Degree $\sbp$ Lower Bounds]\label{thm:small-deg-VP}
Let $d\leq n$ be growing parameters satisfying $d = \omega(\log n)$. There is a $\Theta(dn)$-variate degree $d$ set-multilinear polynomial $F_{n,d}$ in $\VP$ such that $F_{n,d}$  cannot be computed by a $\sbp$ of total-width $\poly(n)$.
\end{theorem}

Next, we formally state the aforementioned hardness escalation result of \cite{BDS23}. In words, {in order to show $\VBP\neq \VP$, it suffices to show lower bounds for any $\sbp$ computing a polynomial $P$ whose degree is at most \emph{about logarithmic} in the number of variables.} 
Towards this goal,  our main result above (Theorem \ref{thm:small-deg-VP}) shows a super-polynomial lower bound for any $\sbp$ computing an explicit set-multilinear polynomial, whose degree is \emph{barely super-logarithmic} in the number of variables. In this sense, it approaches the resolution of Valiant's conjecture.

\begin{theorem}[Hardness Escalation of \cite{BDS23}]\label{thm:hard-esc}
Let $n, d$ be growing parameters with $d = O(\log n/\log \log n)$. Let $P_{n,d}$ be a $\Theta(dn)$-variate degree $d$ set-multilinear polynomial in $\VP$ (respectively, $\VNP$). If $P_{n,d}$ cannot be computed by a $\sbp$ of total-width $\poly(n)$, then
$\VBP\neq \VP$ (respectively, $\VBP \neq \VNP$).
\end{theorem}

Next, we also give an explicit set-multilinear polynomial (with polynomially-large degree) such that any $\sbp$ computing it must require \emph{exponential} total-width. 
This strongly answers a question left open in both \cite{ArvindR16} and \cite{BDS23}. 


\begin{theorem}[Exponential Lower Bounds for $\sbp$]\label{thm:large-deg-VP}
There is a set-multilinear polynomial $F_{n,n}$ in $\VP$, in  $\Theta(n^2)$ variables and of degree $\Theta(n)$, such that any $\sbp$ computing $F_{n,n}$ requires total-width ${\exp(\Omega(n^{1/3}))}$.
\end{theorem}

\autoref{thm:large-deg-VP} and \autoref{thm:small-deg-VP} are also true when $F_{n,d}$ (as defined in Section \ref{sec:VP}) is replaced by the appropriate Nisan-Wigderson polynomial $NW_{n,d}$ (as defined in Section \ref{subsec:NW}), which is known to be in $\VNP$. In fact, we first indeed established them for the Nisan-Wigderson polynomial, and then used some of the ideas presented in a recent work by Kush and Saraf (\cite{KushS23}) to make the hard polynomial lie in $\VP$. {We also acknowledge that an exponential lower bound  -- with weaker quantitative parameters -- for the related model of multilinear ROABPs was obtained in \cite{RR20}. For a  comparison of this model with the $\sbp$ model, see Sections \ref{subsec:roabps} and \ref{sec:RaghusResults}.  
}

With additional effort, and building upon the machinery\footnote{This is explained in more detail in Section \ref{subsec:overview}.} of \cite{KushS23} (which, in turn, uses the techniques developed in \cite{DvirMPY12}), we can almost recover the same lower bounds as in \autoref{thm:large-deg-VP} and \autoref{thm:small-deg-VP} for a set-multilinear polynomial even in $\VBP$. We preferred to first state \autoref{thm:large-deg-VP} and \autoref{thm:small-deg-VP} in the manner above because (i) the proof is  less intricate and in fact, even serves as a prelude to the proof of the latter, and (ii) to draw a direct comparison and contrast with the hardness escalation statement (\autoref{thm:hard-esc}). We now state these results for when the hard polynomial is the $\VBP$ polynomial and then describe two intriguing consequences.

\begin{manualtheorem}{\ref{thm:large-deg-VP}'}\label{thm:large-deg-VBP}
    There is a fixed constant $\delta \geq 1/100$ and a set-multilinear polynomial $G_{n,n}$ in $\VBP$, in  $\Theta(n^2)$ variables and of degree $\Theta(n)$,  such that any $\sbp$ computing $G_{n,n}$ requires total-width ${\exp(\Omega(n^{\delta}))}$.
\end{manualtheorem}

\begin{manualtheorem}{\ref{thm:small-deg-VP}'}\label{thm:small-deg-VBP}
Let $d\leq n$ be growing parameters satisfying $d = \omega(\log n)$. There is a  $\Theta(dn)$-variate, degree $\Theta(d)$ set-multilinear polynomial $G_{n,d}$ in $\VBP$ such that $G_{n,d}$  cannot be computed by a $\sbp$ of total-width $\poly(n)$.
\end{manualtheorem}

The first intriguing consequence of proving the statements above is that we are able to show that the ABP set-multilinearization process given in \cite{BDS23} is nearly tight, as $G_{n,d}$ is known to have a small set-multilinear branching program and yet, any $\sbp$ computing it must have large total-width. To make this point effectively, we first state the following key ingredient in the proof of \autoref{thm:hard-esc}, and subsequently state our tightness result.

\begin{lem}[ABP Set-Multilinearization in \cite{BDS23}]\label{lem:ABP-sml}
Let $P_{n,d}$ be a polynomial of degree
$d$ that is set-multilinear with respect to the partition $X = (X_1,\ldots,X_d)$ where $|X_i| \leq n$ for all $i \in [d]$. If $P_{n,d}$ can be computed by an ABP of size $s$, then it can also be computed by a $\sbp$ of max-width $s$ and total-width $2^{O(d\log d)} s$.   
\end{lem}

\begin{theorem}[Near-Tightness of ABP Set-Multilinearization]\label{thm:tightness}
    For large enough integers $\omega(\log n) = d\leq n$, there is a polynomial $G_{n,d}(X)$ which is set-multilinear over the variable partition $X = (X_1,\ldots,X_d)$ with
each $|X_i| \leq n$, and such that:
\begin{itemize}
    \item it has a branching program of size $\poly(n)$,
    \item but any $\sbp$ of max-width $\poly(n)$ computing $G_{n,d}$ requires total-width $2^{\Omega(d)}$. 
\end{itemize}
\end{theorem}

The second intriguing consequence is the fact that \autoref{thm:tightness} can also be viewed as an exponential separation between the model of (general) small-width set-multilinear branching programs and the model of sums of small-width ordered set-multilinear branching programs. Moreover, we can improve this bound much further in the case of a single ordered set-multilinear branching program.
More precisely, in \autoref{thm:sep-ord-unord} below, we answer a question posed in \cite{KushS23} about the relative strength of an unordered and (a single) ordered set-multilinear branching program, by obtaining a \emph{near-optimal} separation. 
A priori, as is shown in \cite{KushS23} and as mentioned earlier in the introduction, if these two models coincided (i.e., if a general set-multilinear ABP could be simulated by a small and ordered one), then it would have led to super-polynomial lower bounds for general algebraic formulas. 

\begin{theorem}[Near-Optimal Separation between Ordered and Unordered smABPs]\label{thm:sep-ord-unord}
    There is a polynomial $G_{n,d}(X)$ which is set-multilinear over the variable partition $X = (X_1,\ldots,X_d)$ with
each $|X_i| \leq n$, and such that:
\begin{itemize}
    \item it has a set-multilinear branching program of size $\poly(n,d)$,
    \item but any ordered set-multilinear branching program computing $G_{n,d}$ requires width $n^{\Omega(d)}$. 
\end{itemize}
\end{theorem}

Note that $G_{n,d}$ has at most $n^{d}$ monomials and so, it trivially has an ordered set-multilinear ABP of width $n^{d}$. Therefore, the lower bound above is essentially optimal.

\subsection{The ROABP Perspective}\label{subsec:roabps}

One can also view all of our results described in Section \ref{subsec:our-results} through the lens of another well-studied model in algebraic complexity theory, namely \emph{read-once oblivious algebraic branching programs} (ROABPs). 

\begin{definition}[ROABP]\label{def:ROABP}
For integers $n,d$ and a permutation $\sigma \in S_n$, an ABP over the variables $x_1,\ldots,x_n$ is said be a read-once oblivious algebraic branching program (ROABP) in the order $\sigma$ of individual degree $d$ if for each $\ell\in [n]$, all edges from layer $\ell - 1$ to $\ell$ are labelled by univariate polynomials in $x_{\sigma(i)}$ of degree at most $d$.
\end{definition}

ROABPs were first introduced in this form by Forbes and Shpilka in \cite{ForbesS13}, where it is also noted that proving lower bounds against ordered set-multilinear ABPs (as in Definition \ref{def:smABP}) is equivalent to proving lower bounds against ROABPs as well as non-commutative ABPs.

Suppose $f \in \F[X_1, \ldots, X_d]$ is a set-multilinear polynomial with respect to $X_1 \sqcup \cdots \sqcup X_d$ with $X_i = \{x_{i,1}, \ldots, x_{i,n}\}$.
Then we can define an associated polynomial $g_f \in \F[x_1, \ldots, x_d]$ as follows. 
\[
    g_f(x_1, \ldots x_d) = \sum_{\mathbf{e} \in [n]^d} \prod_{i=1}^{n} x_i^{e_i} \cdot \text{ coefficient of } x_{i,e_i}.
\]
Now let us assume that $g_f$ can be computed by an ROABP of size $s$ that is ordered with respect to $\sigma \in S_n$.
Then a set-multilinear ABP ordered with respect to $\sigma$ can be constructed using it, by simply replacing $x_i^{e_i}$ by $x_{i,e_i}$ and erasing any degree zero components on each edge.
It is easy to check that this computes $f$ and we can use the lower bound against ordered set-multilinear ABPs for $f$ to prove a lower bound against ROABPs for $g_f$.
Conversely, given $g \in \F[x_1, \ldots, x_n]$, we can define $f_g \in \F[X_1, \ldots, X_n]$ with $X_i = \{x_{i,0}, x_{i,1}, \ldots, x_{i,d}\}$ by by replacing $x_i^{e_i}$ with $x_{i,e_i}$.
We could then use an ordered set-multilinear ABP computing $f_g$ to construct an ROABP (in the same order) computing $g$ by using the inverse transformation, thereby proving that lower bounds against ROABPs imply lower bounds against ordered set-multilinear ABPs.
Furthermore, the computation that an ROABP (or an ordered set-multilinear ABP) performs can be seen to be \emph{non-commutative}. 
This is because the variables (or linear forms) along a path get multiplied in the \emph{same} order $\sigma$ as that of the ROABP (or ordered set-multilinear ABP).

As a consequence, exponential lower bounds follow for a single ROABP from the work of Nisan (\cite{Nisan91}), and also from later works (\cite{Jansen08, KayalNS20}). 
Using the  transformation described above, our lower bounds (Theorems \ref{thm:large-deg-VP} and \ref{thm:large-deg-VBP}) can also be viewed as exponential lower bounds for the model of sum of ROABPs. 
The work of Ramya \& Rao \cite{RR20} also prove (weaker) exponential lower bounds against this model for a multilinear polynomial computable by multilinear circuits.
In a follow-up work, Ghoshal \& Rao \cite{GR21} prove an exponential lower bound against sums of ROABPs with the additional restriction that the summand ROABPs have pollynomially-bounded width for a mulilinear polynomial computable by multlinear ABPs.
On the other hand, the works of Arvind \& Raja (\cite{ArvindR16}) and Bhargav, Dwivedi \& Saxena (\cite{BDS23} provide lower bounds in certain restricted versions of this model. 
Along with these, the work of Anderson, Forbes, Saptharishi, Shpilka, and Volk (\cite{AndersonFSSV18}) also implies an exponential lower bound for a restricted version (for the sum of $k$ ROABPs when $k = o(\log n)$).

Finally, we note that ROABPs have been studied extensively in the context of another central problem in algebraic complexity theory: that of  polynomial identity testing (PIT). The PIT question for a general algebraic model $\mathcal{M}$ is the following:
Given access to an $n$-variate polynomial $f$ of degree at most $d$ that can be computed in the model $\mathcal{M}$ of (an appropriate measure of) complexity at most $s$, determine whether $f \equiv 0$ in $\poly(n,d,s)$ time.
When one is given access to the model computing $f$ explicitly, this flavour of PIT is called white-box PIT, and when one is merely provided query access to $f$,  it is called black-box PIT. 

The solution to the PIT problem for ROABPs in the white-box setting
follows from a result by Raz and Shpilka (\cite{RazS05} -- where it is stated in the equivalent language of non-commutative computation). However, the corresponding problem in the black-box setting remains open to this date, with the best-known time bound in the black-box setting still being only $s^{O(\log s)}$ due to the work by Forbes and Shpilka (\cite{ForbesS13}), who additionally assumed that the ordering of the ROABP is known. This was matched later by Agrawal Gurjar, Korwar, and Saxena (\cite{AgrawalGKS15}) in the unknown order setting, improving upon the work of Forbes, Saptharishi and Shpilka (\cite{ForbesSS14}). Guo and Gurjar improved the result further by improving the dependence on the width \cite{guo2020improved}. 
Additionally, there have been various improvements to this result in restricted settings (\cite{GurjarKS17, GurjarV20, BhargavaG22}) and some other works that study PIT for a small sum of ROABPs (\cite{GurjarKST17, BishtS21,guo2020improved}). When the number of summands is super-constant, the question of even white-box PIT remains wide open.

\subsection{Related Work}\label{sec:RaghusResults}

In this subsection, we discuss two closely related papers, namely those of Ramya \& Rao \cite{RR20} and Ghoshal \& Rao \cite{GR21}\footnote{We thank Ben Lee Volk and Utsab Ghosal for pointing out these papers to us after the release of an initial pre-print of this article, which erroneously claimed that it was the first to show super-polynomial lower bounds in the sum of ROABPs model.}, which study the model of sum of ROABPs in the multilinear setting.
In \cite[Theorem 1]{RR20}, the authors show that there exists an explicit multilinear polynomial (computable by a small multilinear circuit) such that any sum of ROABPs computing it has exponential size. 
In \cite[Theorem 2]{GR21}, the authors show a similar lower bound for an explicit multilinear polynomial (computable by a small multilinear ABP) -- albeit, in the restricted setting where the summand ROABPs have polynomially-bounded width.

Using the transformation described in Section \ref{subsec:roabps}, one can then view these lower bounds as ones against the $\sbp$ model in the special case that each bucket in the variable partition has size $2$. (To see how a multilinear polynomial say over the variable set $x_1,\ldots,x_d$ can be set-multilinearized trivially, here is a sketch: for each variable $x_i$, have a variable set $X_i$ comprising of two fresh variables $x_{i,0}$ and $x_{i,1}$ in the new set-multilinear polynomial; here, the latter is to signify the ``presence" of $x_i$ in any monomial of the original multilinear polynomial, whereas the former is to signify its ``absence".)
Additionally, it is not hard to see that the set-multilinearized version of the hard polynomials (in the manner just described) used in \cite[Theorem 1]{RR20} and \cite[Theorem 2]{GR21} are efficiently computable by small set-multilinear circuits and set-multilinear ABPs respectively.
We note, however, that even so, our result in the high-degree setting where the hard polynomial is in $\VP$ (\autoref{thm:large-deg-VP}) is quantitatively better than \cite[Theorem 1]{RR20}.
Additionally, our result in the high-degree setting where the hard polynomial is in $\VBP$ (\autoref{thm:large-deg-VBP}) is \emph{both} quantitatively as well as qualitatively better than \cite[Theorem 2]{GR21} -- the latter since we do not assume any bound on the width of the individual summand ordered set-multilinear ABPs.
More crucially, 
our techniques enable us to prove super-polynomial bounds even when the degree is vastly smaller than the number of variables -- in particular, when $d$ is as low as $\omega(\log n)$ (\autoref{thm:small-deg-VP} and \autoref{thm:small-deg-VBP}) -- which is the more interesting regime of parameters due to the work of \cite{BDS23}. 

Ramya and Rao \cite{RR20} also study another model, which they call \emph{sum of $\alpha$-set-multilinear ABPs}. 
They define $\alpha$-set-multilinear ABPs to be ABPs with $N^{\alpha}$ layers, where $N$ is the number of variables in the polynomial being computed.
Any edge between layer $\ell-1$ and $\ell$ in an $\alpha$-set-multilinear ABP is labelled by an arbitrary multilinear polynomial over $X_\ell$, where $X = X_1 \sqcup \cdots \sqcup X_{N^\alpha}$ is a partition of the variable set. 
Then, for $\alpha \geq 1/10$, they establish exponential lower bounds against sum of $\alpha$-set-multilinear ABPs for a polynomial that is multilinear, but which is \emph{not} set-multilinear under the variable partition that the model respects.
Hence, even though this model is more general than ordered set-multilinear ABPs, this result \cite[Theorem 3]{RR20} is also not comparable with ours as
our hard polynomial is set-multilinear.
Again, more crucially, the result \cite[Theorem 3]{RR20} does not handle the ``low-degree'' regime --- a setting in which our techniques allow us to prove lower bounds.

\subsection{Proof Overview}\label{subsec:overview}

The organization of this subsection is as follows: we first describe the basics of the partial derivative method and summarize its typical application in proving lower bounds against a generic set-multilinear model of computation. Next, we briefly describe Nisan's original partial derivative method from \cite{Nisan91} to prove lower bounds specifically against a single ordered set-multilinear branching program. We then describe an alternative approach that yields a slightly weaker bound for the same model, but nevertheless is versatile enough that we can generalize it considerably more in order to prove Theorems \ref{thm:large-deg-VP} and \ref{thm:small-deg-VP}. Finally, we describe the additional ideas needed in order to situate the hard polynomial in these theorems in $\VBP$ and in the process, establish the tightness result for ABP set-multilinearization (Theorem \ref{thm:tightness}).


\paragraph{Partial Derivative Measure Basics:}
The high-level idea is to work with a {measure} that we show to be ``small'' for all polynomials computed by a specified model of computation -- the model against which we wish to prove lower bounds. If we can also show that there is a ``hard'' polynomial for which the measure is in fact ``large'', then it follows that this polynomial cannot be computed by the specified model. These \emph{partial derivative measures}, after the initial work (\cite{Nisan91}) by Nisan, were further developed by 
Nisan and Wigderson in \cite{NisanW97}, who used them to prove some constant-depth set-multilinear formula lower bounds. Since then, variations of these measures have also been used to prove various other stronger set-multilinear {formula} lower bounds (e.g., \cite{LST1, LST2, LST3, mfcs/BhargavDS22, KushS22, KushS23}).

Given a variable partition $(X_1,\ldots, X_d)$, the idea is to label each set of variables $X_i$ as `$+1$' or `$-1$' according to \emph{some} rule (called a ``word") $w\in \bits^d$. Let $\calP_w$ and $\calN_w$ denote the set of positive and negative indices (or coordinates) respectively, and let $\calM_w^\calP$ and $\calM_w^\calN$ denote the sets of all set-multilinear monomials over $\calP_w$ and $\calN_w$ respectively. For a polynomial $f$ that is set-multilinear over the given variable partition $(X_1,\ldots, X_d)$, the measure then is simply the rank of the ``partial derivative matrix'' $\calM_w(f)$, whose rows are indexed by the elements
of $\calM_w^{\calP}$ and columns indexed by $\calM_w^{\calN}$, and the entry of this matrix corresponding to a
row $m_1$ and a column $m_2$ is the coefficient of the monomial $m_1\cdot m_2$ in $f$. 

For a subset $S\subseteq[d]$, let $w_S$ denote the sum of those coordinates of $w$ that lie in $S$. In other words, $|w_S|$ measures the amount of ``bias" that the rule $w$ exhibits when restricted to the $S$ coordinates. Note that the rank of $M_w(f)$ can never exceed $n^{(d-|w_{[d]}|)/{2}}$. Furthermore, we have that the rank measure is \emph{multiplicative}: if $f$ and $g$ are polynomials that are set-multilinear over \emph{disjoint} subsets of the global partition $(X_1,\ldots, X_d)$, then the rank of $\calM_w(f\cdot g)$ is the product of the ranks of $\calM_w(f)$ and $\calM_w(g)$. These two observations, combined with the sub-additivity of rank, provide a recipe for showing lower bounds against any given set-multilinear model of computation: the overall idea is to carefully split up the original model into smaller, multiplicatively disjoint parts and then argue the existence of a rule for which enough of these parts exhibit high bias. This process allows us to prove that the measure is small for the model of computation. Therefore, one can conclude that any explicit polynomial for which the measure is provably high -- which needs to established separately -- can not be computed by this model.  
It is known (\cite{KushS22,KushS23}) that there is a set-multilinear polynomial 
$NW_{n,d}$ in $\VNP$ (see Section \ref{subsec:NW}) as well as a set-multilinear polynomial $F_{n,d}$ in $\VP$ (see Section \ref{sec:VP}) for which the matrices $\calM_w(NW_{n,d}),\calM_w(F_{n,d}), $ have full-rank, whenever $|\calP_w| = |\calN_w|$.

\paragraph{Nisan's original lower bound:}Let us first summarize how Nisan's original partial derivative method from \cite{Nisan91}, as alluded to in Section \ref{subsec:sml}, can be applied in this context to obtain lower bounds against the size of a single ordered set-multilinear ABP (ordered smABP) computing the aforementioned ``full-rank'' polynomials. Given any set-multilinear branching program $A$ ordered with respect to some permutation $\sigma\in S_d$ computing $F_{n,d}$, the idea is to pick a word $w$ such that the $+1$ labels in $w$ precisely correspond to the ``left half" of the ordering $\sigma$, and the $-1$ labels correspond to the ``right half". One can then observe that the rank of $\calM_w(F_{n,d}) = \calM_w(A)$ serves as a lower bound on the number of nodes $s$ in the middle layer of the ABP, yielding a near-optimal $n^{\Omega(d)}$ lower bound: this is because the matrix $\calM_w(A)$ is easily seen to be the product of an $n^{d/2}\times s$ and an $s\times n^{d/2}$ matrix.

We now sketch an alternate proof: rather than constructing a word dependent on the ordering of variable sets $X_i$ in the ordered smABP $A$ as above, choose a uniformly random\footnote{\label{fn:condition}We also need to suitably condition on 
the event that the word $w$ is symmetric (i.e., $|\calP_w| = |\calN_w|$) in order to use the full-rank property of the hard polynomial -- the probability of this event is $\Theta(\frac{1}{\sqrt{d}})$. For ease of exposition, we omit the technical details in this sketch.} word $w$ from $\bits^d$. We demonstrate that, with positive probability, the rank of $\calM_w(A)$ is bounded by $s \cdot n^{d/2-\Omega(\sqrt{d})}$, where $s$ is the width of the middle layer in $A$: Standard anti-concentration bounds imply that, with at least constant probability, the bias in the left and right halves of $A$ is $\Omega(\sqrt{d})$. Since $A$ can be expressed as a sum of $s$ polynomials $f_i\cdot g_i$ for $i\in[s]$, where each $f_i$ and $g_i$ are ordered smABPs with respect to disjoint subsets of the global partition, we encounter a loss of a factor of $n^{\Omega(\sqrt{d})}$ in the rank of the product polynomial $\calM_w(f_i\cdot g_i)$ due to the bias of $w$. This, combined with the sub-additivity of rank, shows the desired bound of $s \cdot n^{d/2-\Omega(\sqrt{d})}$ on the rank of $\calM_w(A)$. Finally, we exploit the full-rank property of $F_{n,d}$ with respect to such words to establish a lower bound of $n^{\Omega(\sqrt{d})}$ on the width $s$ of a single ordered smABP computing $F_{n,d}$. Notably, this bound is indeed slightly worse than what one can obtain by manually defining a rule $w$ deterministically, which ensures a \emph{maximal} bias of $d/2$ in each half of $A$ as described in the paragraph above.

\paragraph{Generalization of the alternative argument:} 
The alternative argument described above yields an exponential lower bound even for a \emph{sum} of ordered smABPs, assuming the number of summands is small. Consider a $\sbp$ of the form $\sum_{i=1}^t A_i$, of max-width $s$, computing $F_{n,d}$. For each summand $A_i$, the analysis above provides an upper bound of $s \cdot n^{d/2-\Omega(\sqrt{d})}$ on the rank of $\calM_w(A_i)$ with constant probability. If the number of summands $t$ is a small enough constant, the union bound ensures the existence of a word $w$ such that the rank of $\calM_w(\sum A_i)$ is at most $t\cdot s\cdot n^{d/2-\Omega(\sqrt{d})}$. Thus\footnote{See footnote \ref{fn:condition}.}, we obtain an exponential lower bound on $t\cdot s$ since this $\sbp$ computes a full-rank polynomial. However, because of the use of the union bound in this manner, this method faces an inherent limitation -- it is unable to handle more than a very small number of summands, even if we lower the bias demand from each half (e.g., from $\Omega(\sqrt{d})$ to $\Omega(\sqrt[4]{d})$ or a smaller polynomial in $d$). In fact, one can construct a sum of $d$ ordered smABPs (by starting with a single smABP ordered arbitrarily and considering the $d$ cyclic shifts of this ordering) such that any unbiased word $w$ (i.e., $w_{[d]} = 0$) has the property that for at least one of the summands, the left and right halves will have no bias! Evidently then, in order to prove lower bounds against an \emph{unrestricted} number of summands, we need another method to analyze the rank of the summands. Nonetheless, a conceptual takeaway from the exercise above is that selecting a rule $w$ that is oblivious to the orderings of individual summands (and in particular, a \emph{random} rule) still lets us derive strong lower bounds for the sum of \emph{multiple} ordered smABPs.

Suppose instead of slicing an ordered smABP $A$ down the middle, we slice it into three roughly equal pieces. Then, it is possible to write the polynomial computed by $A$ as a sum over $s^2$ terms, each of the form $f_i\cdot g_i\cdot h_i$ where for each $i$, each of $f_i, g_i, h_i$ depends on $d/3$ \emph{disjoint} variable sets of the global partition. We can then perform a similar analysis as above to show enough bias across these $3$ pieces, thereby obtaining a rank deficit. More precisely, we can conclude that for a single ordered smABP $A$, again with a constant probability, the rank of $\calM_w(A)$ is at most $s^2 \cdot n^{d/2-\Omega(\sqrt{d})}$. When we slice the ABP into $3$ pieces in this way, it is not immediately clear where the gain is. In fact, for a single ordered smABP, this method actually gives a worse lower bound on $s$ due to the presence of the factor of $s^2$. Where we gain is in the magnitude of the \emph{probability} with which we can guarantee that a single ordered smABP has a rank deficit -- we will now describe how this observation allows us to take a union bound over many more summands.

In order to illustrate this trade-off more clearly, we will partition the ordered smABP $A$ into many more pieces. Suppose we slice it into $q \approx \sqrt{d}$ pieces, each of size roughly $r = d/q \approx \sqrt{d}$ (this is just one setting of parameters;
$q$ and $r$ are suitably optimized in the final proof).
Thus, the polynomial that $A$ computes can be written as a sum of at most $s^{q-1}$ terms, where each term is a product of $q$ polynomials -- each set-multilinear over a disjoint subset of the global partition, where each piece has size $r$. When a word $w$ is chosen randomly, each such piece again exhibits a bias of  about $\Omega(\sqrt{r})$ with constant probability. The crucial observation then is that by known concentration bounds, it can be shown that with
probability \emph{exponentially} close to $1$, the sum of the biases across all the $q$ pieces is $\Omega(q\sqrt{r}) = \poly(d)$.
For a single ordered smABP $A$, this shows that the rank of $\calM_w(A)$ is at most $s^{q} \cdot n^{-\Omega(q\sqrt{r})}$, which is still enough to show an exponential lower bound on $s$, even though it is worse
than what we obtained by slicing into fewer pieces. 

The key advantage in implementing this analysis is that it provides a way to argue that for a random word $w$, $\calM_w(A)$ has low rank for a single ordered smABP $A$ -- with probability \emph{exponentially} close to $1$.
In particular, this allows us to union bound over exponentially many ordered smABPs and show that even if we have an $\sbp$ computing $F_{n,d}$ of exponential support size, with high probability, each summand will have a rank deficit. Then, again using the sub-additivity of rank, we can conclude that the sum has a rank deficit as well.

This method of analyzing the rank of an ordered smABP by partitioning it into {\em numerous} pieces and tactfully using concentration bounds is novel, and conceptually the most essential aspect of the proof. As we demonstrated above, this method of analysis indeed gives a worse bound for a single smABP. However, while mildly sacrificing what we can prove about the rank of a single ordered smABP, we are able to leverage it to still prove something meaningful about the rank of a {\it sum} with a much larger number of summands. 

Our partial derivative measure draws inspiration from previously known lower bounds in the context of multilinear and set-multilinear {\it formulas~}(\cite{Raz06, KushS22}). One noteworthy distinction lies in the analysis of the measure: whereas the partitioning is present intrinsically in those formula settings, in our setting of ABPs, we deliberately introduce the partitioning at the expense of a notable increase in the number of summands or the total-width (and therefore, in the number of events we union bound over). The substantial advantage gained in utilizing this partitioning for rank analysis justifies the tolerable increase in the total-width.


\paragraph{Tightness of ABP set-multilinearization:}
In order to make the hard polynomial in Theorems \ref{thm:large-deg-VP} and \ref{thm:small-deg-VP} lie in $\VBP$, one might wonder if we can get away with using the same rank measure (i.e., rank of the matrix $M_w(\cdot)$ for a uniformly random word $w\in\bits^d$) that was used in the analysis above for the $\VP$ polynomial $F_{n,d}$. However, as far as we know, full-rank polynomials (in the sense described above) may also require super-polynomial sized set-multilinear ABPs. Thus, in order to prove a separation between (general)
set-multilinear ABPs and (sums of) ordered set-multilinear ABPs, we seek a property that is weaker than being full-rank and yet is still
useful enough for proving lower bounds against our model. For this, we rely upon the \emph{arc-partition} framework that is developed in \cite{KushS23} in order to prove near-optimal set-multilinear formula lower bounds (building upon the initial ingenious construction given in \cite{DvirMPY12} for the multilinear context), tailor the framework to our $\sbp$ model, and use a more delicate concentration bound analysis in order to prove our results.

An \emph{arc-partition} is a special kind of symmetric word $w$  from $\bits^d$: we will now describe a distribution over $\bits^d$; the words that will
have positive probability of being obtained in this distribution will be called arc-partitions. The
distribution is defined according to the following (iterative) sampling algorithm. Position the $d$
variable sets on a cycle with $d$ nodes so that there is an edge between $i$ and $i+ 1$ modulo $d$. Start with
the arc $[L_1, R_1] = \{1, 2\}$ (an arc is a connected path on the cycle). At step $t > 1$ of the process,
maintain a partition of the arc $[L_t
, R_t
]$. “Grow” this partition by first picking a pair uniformly
at random out of the three possible pairs $\{L_t - 2, L_t - 1\},\{L_t - 1, R_t + 1\},\{R_t + 1, R_t + 2\}$, and then choosing a labelling (or partition) $\Pi$ on this pair i.e., assigning one of them `$+1$' and the other `$-1$' uniformly at random. After $d/2$ steps, we have chosen a partition (i.e., a word $w$ from $\bits^d$) of the $d$ variable sets into two disjoint, equal-size
sets of variables $\calP$ and $\calN$. It is known from \cite{KushS23} that there exist set-multilinear polynomials $G_{n,d}$ (as defined in Section \ref{subsec:VBP}) that are \emph{arc-full-rank} i.e., $\calM_w(G_{n,d})$ is full-rank for every arc-partition $w$. Analogous to the proofs of Theorems \ref{thm:large-deg-VP} and \ref{thm:small-deg-VP}, we establish our $\sbp$ lower bounds by showing that with high probability, every $\sbp$ has an appropriately large rank deficit with respect to the arc-partition distribution.  However, as we now briefly explain, this analysis turns out to be significantly more intricate.

Similar to the analysis as in the $\VP$ case, we partition an ordered smABP $A$ into $q$ pieces of size $r$ each, and write the polynomial that it computes as a sum of at most $s^q$ terms. Again, the task is to show that an arc-partition $w$ exhibits a large total bias across the $q$ pieces: more precisely, we show that if the pieces are labelled as $S_1,\ldots,S_q$, then with probability \emph{exponentially} close to $1$, the sum $\sum_{i = 1}^q |w_{S_i}|$ (i.e., the total bias of $w$ across these pieces) is $\Omega (q r^\eps )$, which is polynomially large in $d$ for an appropriate setting of $q,r$. This then yields the desired rank deficit similar to the $\VP$ analysis (albeit with mildly worse parameters).

The bias lower bound is established in the following sequence of steps:
\begin{itemize}
    \item View the partition $(S_1,\ldots,S_q)$ of $[d]$ as a fixed ``coloring'' of the latter. We say that a \emph{pair} -- as sampled in the construction of an arc-partition described above -- ``violates'' a color $S$ if exactly one of the elements of the pair is colored by the set $S$. Then, we show that with probability exponentially close to $1$, ``many'' colors must have ``many'' violations: more precisely, that at least a constant fraction of the colors (i.e., $\Omega(q)$ many) have at least $r^{2\eps}$ many violations each (for some small constant $\eps>0$). Such a ``many violations'' lemma is also established in \cite{KushS23} in the context of proving set-multilinear formula lower bounds. We show that this lemma, in fact, holds for a much wider range of parameters than was previously known; this extension is indeed necessary for our use. The proof of this strengthened many violations lemma is deferred to the appendix.
    \item We then use the strengthened many violations lemma to argue that even though $w$ is not chosen uniformly at random and as such, its coordinates are not truly independent, it possesses ``enough'' inherent independence  that a similar concentration bound as in the $\VP$ analysis is applicable. More precisely, we show that with high probability, there is an ordering of a set of $\Omega(q)$ colors such that each such color has at least $r^{2\eps}$ violations and a more nuanced application of standard concentration bounds shows that $w$ exhibits a total bias of at least $\Omega(qr^\eps )$.
\end{itemize}

\subsection{Structure of the paper}
In Section \ref{sec:prel}, we formally introduce the measure used to prove our lower bounds, as well as the hard polynomials for which the bounds are shown.
Section \ref{sec:VP} is dedicated to proving lower bounds against $\sbp$ in both the high degree and low degree regimes for a polynomial in $\VP$ (\autoref{thm:large-deg-VP}, \autoref{thm:small-deg-VP}).
We then prove the corresponding results for a polynomial in $\VBP$ (\autoref{thm:large-deg-VBP}, \autoref{thm:small-deg-VBP}), in Section \ref{sec:tightness}, thereby proving a near-tightness of the hardness escalation result in \cite{BDS23} (\autoref{thm:tightness}).
Finally, in Section \ref{sec:opt-sep-sml-osml}, we show an optimal separation between setmultilinear and ordered set-multilinear ABPs (\autoref{thm:sep-ord-unord}).

\section{Preliminaries}\label{sec:prel}
\subsection{Relative Rank and its Properties}

We first describe the notation that we need to define the measures that we use to prove our results described in Section \ref{subsec:our-results}. Instead of directly working with the rank of the partial derivative matrix, we work with the following normalized form.

\begin{definition}\label{def:rk-rest}
Let $w = (w_1, w_2,\ldots, w_d)$ be a tuple (or word) of non-zero real numbers. For a subset $S \subseteq [t]$, we shall refer to the sum $\sum_{i\in S} w_i$ by $w_S$,  and by $w|_S$, we will refer to the tuple obtained by considering only the
elements of $w$ that are indexed by $S$. 
Given a word $w=(w_1,\ldots,w_d)$, we denote by $\ol{X}(w)$ a tuple of $d$ sets of variables $(X(w_1),\ldots,X(w_d))$ where
$|X(w_i)| = 2^{|w_i|}$.\footnote{In particular, $2^{|w_i|}\in\mathbb{N}$.}
We denote by $\Fsm[\calT]$ the set of set-multilinear polynomials over the tuple
of sets of variables $\calT$.
\end{definition}

\begin{definition}[Relative Rank Measure of \cite{LST1}]\label{def:relrk}
Let $\ol{X}=(X_1,\ldots,X_d)$ be a tuple of sets of variables such that $|X_i|=n_i$ and let $f\in \Fsm[\ol{X}]$.
 Let $w = (w_1, w_2,\ldots, w_d)$ be a tuple (or word) of non-zero real numbers such that $2^{|w_i|}  = n_i $ for all $i \in [d]$. Corresponding to a word $w$, define $\calP_w \coloneqq \{i\ |\ w_i > 0\}$ and $\calN_w \coloneqq \{i\ |\ w_i < 0\}$. Let $\calM^{\calP}_{w}$ be the
set of all set-multilinear monomials over the subset of the variable sets $X_1, X_2,\ldots, X_d$ precisely
indexed by $\calP_w$, and similarly let $\calM^{\calN}_{w}$ be the set of all set-multilinear monomials over these variable
sets indexed by $\calN_w$.

Define the ‘partial derivative matrix’ matrix $\calM_w(f)$ whose rows are indexed by the elements
of $\calM^{\calP}_w$ and columns indexed by the elements of $\calM^{\calN}_{w}$ as follows: the entry of this matrix corresponding to a
row $m_1$ and a column $m_2$ is the coefficient of the monomial $m_1\cdot m_2$ in $f$. We define
\[
\rk_w(f) \coloneqq \frac{\mathrm{rank}(\calM_w(f))}{\sqrt{|\calM^{\calP}_w|\cdot |\calM^{\calN}_w|}} = \frac{\mathrm{rank}(\calM_w(f))}{2^{\frac{1}{2}\sum_{i\in[d]}|w_i|}}.
\]
\end{definition}

The following is a simple result that establishes various useful properties of the relative rank measure.
\begin{claim}[\cite{LST1}]\label{clm:rk-props}
\begin{enumerate}
    \item(Imbalance) Say $f \in \Fsm[\ol{X}(w)]$. Then, $\rk_w(f)\leq 2^{-|w_{[d]}|/2}$.
    \item(Sub-additivity) If $f,g \in \Fsm[\ol{X}(w)]$, then $\rk_w(f+g)\leq \rk_w(f)+\rk_w(g)$.
    \item(Multiplicativity) Say $f = f_1 f_2\cdots f_t $ and assume that for each $i\in [t]$, $f_i\in \Fsm[\ol{X}(w|_{S_i})]$, where $(S_1, \ldots, S_t)$ is a partition of $[d]$. Then
    \[
    \rk_w(f) = \prod_{i\in[t]} \rk_{w|_{S_i}}(f_i).
    \]
\end{enumerate}
\end{claim}

We will repeatedly make use of the following.

\begin{theorem}[Chernoff Bound, as stated in \cite{mitzenmacher_upfal_2005}]\label{thm:Chernoff}
Suppose $X_1, \ldots, X_n$ are independent random variables taking values in $\zo$. Let $X$ denote their sum and let $\mu = \E[X]$ denote the expected value of the sum. Then for any $\delta > 0$,
\begin{eqnarray*} \Pr[X\ge (1+\delta)\mu] &\le& \exp\left(-\frac{\delta^2\mu}{2+\delta}\right), \text{ if }0\leq \delta, \\
  \Pr[X\le (1-\delta)\mu] &\le& \exp\left(-\frac{\delta^2\mu}{2}\right), \text{ if }0< \delta <1, \\ 
  \Pr[|X-\mu|\geq \delta \mu]&\leq& 2\cdot \exp\left(-\frac{\delta^2\mu}{3}\right),\text{ if }0< \delta <1.\\
\end{eqnarray*}
\end{theorem}

\subsection{Inner Product Gadget}
The following observation is used crucially to construct the hard polynomials in $\VP$ as well as $\VBP$. 
\begin{obs}[\cite{KushS23}]\label{obs:ip-full-rank}
    Let $n = 2^k$ and $X_1 = \{x_{1,1},\ldots,x_{1,n}\}$ and $X_2 = \{x_{2,1},\ldots,x_{2,n}\}$ be two disjoint sets of variables. Then, for any \emph{symmetric} word $w\in\{k,-k\}^2$ (i.e., where $w_1 + w_2 = 0$) and for the inner product `gadget' $ f = X_1\cdot X_2 = \sumi x_{1,i}x_{2,i}$, $\rk_w(f) = 1$ i.e., $\calM_w(f)$ is full-rank.
\end{obs}

\subsection{A Hard Set-multilinear Polynomial in $\VNP$}\label{subsec:NW}
As is done in previous lower bounds using the NW polynomials (for example, see \cite{KayalSS14}), we will identify the set of the first $n$ integers as elements of $\F_n$ via an arbitrary correspondence $\phi : [n] \rightarrow \F_n$. If $f(z) \in \F_n[z]$ is a univariate polynomial, then we abuse notation to let $f(i)$ denote the evaluation
of $f$ at the $i$-th field element via the above correspondence i.e., $f(i)\coloneqq \phi^{-1}
(f(\phi(i)))$. To simplify the exposition, in the following definition, we will omit the correspondence $\phi$ and identify a variable
$x_{i,j}$ by the point $(\phi(i), \phi(j)) \in \F_n \times \F_n$.

\begin{definition}[Nisan-Wigderson Polynomials]\label{def:NW}
For a prime power $n$,
let $\F_n$ be a field of size $n$. For an integer $d\leq n$ and the set $X$ of $nd$ variables
$\{x_{i,j} : i\in[n], j \in [d]\}$, we define the degree $d$
homogeneous polynomial $NW_{n,d}$ over any field as
\[
NW_{n,d}(X) = \sum_{\substack{f(z)\in\F_n[z]
\\ \deg(f)<d/2}} \prod_{j\in[d]} x_{f(j),j}.
\]
\end{definition}

\begin{claim}[\cite{KushS22}]\label{clm:NW-full-rk}
For an integer $n = 2^k$ and $d\leq n$, let $w \in \{k,-k\}^d$ with $w_{[d]} = 0$. Then $\rk_w(NW_{n,d}) = 1$ i.e., $\calM_w(NW_{n,d})$ has full rank.
\end{claim}
\begin{proof}
Fix $n = 2^k$ and $d$, so that we can also write $NW$ for $NW_{n,d}$, and let $n' = d/2$. The condition on $w$ implies that $|\calP_w| = |\calN_w| = n'$. Observe that $\calM_w(NW)$ is a square matrix of dimension $|\calM^{\calP}_{w}| = |\calM^{\calN}_{w}| = n^{n'}$. Consider a row of $\calM_w(NW)$ indexed by a monomial $m_1 = x_{i_1,j_1}\cdots x_{i_{n'},j_{n'}}\in \calM^{\calP}_{w}$. $m_1$ can be thought of as a map from $S = \{j_1,\ldots,j_{n'}\}$ to $\F_n$ which sends $j_\ell$ to $i_\ell$ for each $\ell \in [n']$. Next, by interpolating the pairs $(j_1,i_1),\ldots, (j_{n'},i_{n'})$, we know that there exists a unique polynomial $f(z)\in \F_n(z)$ of degree $<n'$ for which $f(j_\ell) = i_\ell$ for each $\ell\in [n']$. As a consequence, there is a unique `extension' of the monomial $x_{i_1,j_1}\cdots x_{i_{n'},j_{n'}}$ that appears as a term in $NW$, which is precisely $m_1\cdot \prod_{j\in \calN_w}x_{f(j),j}$. Therefore,
all but one of the entries in the row corresponding to $m_1$ must be zero, and the remaining entry must be $1$. Applying the same argument to the columns of $\calM_w(NW)$, we deduce that $\calM_w(NW)$ is a permutation matrix, and so has full rank.
\end{proof}

\subsection{A Hard Set-multilinear Polynomial in $\VP$}\label{sec:VP}
Let $d$ be an even integer and let $X = (X_1,\ldots, X_{d})$ be a collection of sets of variables where each $|X_i| = n$, and similarly, let $Y = (Y_1,\ldots, Y_{d})$ be a distinct collection of sets of variables where each $|Y_i| = n$. We shall refer to the $Y$-variables as the \emph{auxiliary} variables. For $i$ and $j \in \{1,\ldots,d\}$, let $X_i \cdot X_j$ denote the inner-product quadratic form $\sum_{k = 1}^n x_{ik}x_{jk}$. Here, we shall assume that $X_i = \{x_{i,1},\ldots, x_{i,n}\}$ and $Y_i = \{y_{i,1},\ldots, y_{i,n}\}$.

For two integers $i \in \N$ and $j \in \N$, we denote $[i, j] = \{k \in \N : i \leq k \text{ and } k \leq j\}$ and call such a set an \emph{interval}. For every interval $[i, j] \subseteq [d]$, we define a polynomial $f_{i,j}(X,Y)\in \Fsm[X_i,\ldots,X_j,Y_i,\ldots, Y_j]$ as follows:

\[
f_{i,j} = \begin{cases}
y_{i,j}y_{j,i}(X_i\cdot X_j) &\text{if $j = i+1$}\\
0 &\text{if $j-i$ is even}\\
y_{i,j}y_{j,i}(X_i\cdot X_j)\cdot f_{i+1,j-1} + \sum_{r = i+1}^{j-1} f_{i,r}f_{r+1,j} &\text{otherwise}\\
\end{cases}
\]

These $f_{i,j}$ in present form were defined in \cite{KushS23}, but were in turn inspired from an earlier work of Raz and Yehudayof (\cite{RazY08}) in the multilinear context. \cite{KushS23} shows that they have the following full-rank property that will be instrumental for us.
\begin{lem}[\cite{KushS23}]\label{lem:Fn-VP-props}
    Let $n = 2^k$ and $d\leq n$ be an even integer. Over any field $\F$ of characteristic zero, the polynomial $F_{n,d} = f_{1,d}\in\Fsm[X,Y]$ as defined above satisfies the following: For any $w\in\{-k,k\}^d$ with $w_{[d]} = 0$, $\calM_w(F_{n,d})$ is full-rank when viewed as a matrix over the field $\F(Y)$, the field of rational functions over the $Y$ variables.
\end{lem}

\subsection{A Hard Set-Multilinear Polynomial in $\VBP$}\label{subsec:VBP}
\subsubsection{Arc-partition Measure Description}
This subsection is adapted from Section 2 of \cite{DvirMPY12}. Let $n = 2^k$, $d\leq n$ be an even integer, and let $X = (X_1, X_2, \ldots, X_{d})$ be a collection of disjoint sets of $n$ variables each.
An \emph{arc-partition} will be a special kind of \emph{symmetric} word $w\in \{-k,k\}^d$ (i.e., a one-to-one map $\Pi$ from $X$ to $\{-k,k\}^d$). For the purpose of this subsection, the reader can even choose to think of the alphabet of $w$ as $\{-1,1\}$ (i.e., one `positive' and one `negative' value) -- we use $k,-k$ only to remain consistent with \autoref{def:relrk}.

Identify $X$ with the set $\{1, 2,\ldots, d\}$ in the natural way. Consider the $d$-cycle
graph, i.e., the graph with nodes $\{1, 2,\ldots, d\}$ and edges between $i$ and $i + 1$ modulo $d$. For
two nodes $i \neq j$ in the $d$-cycle, denote by $[i, j]$ the arc between $i, j$, that is, the set of nodes on the
path $\{i, i + 1, \ldots , j - 1, j\}$ from $i$ to $j$ in $d$-cycle.
First, define a distribution $\calD_P$ on a family of pairings (a list of disjoint pairs of nodes in the
cycle) as follows. A random pairing is constructed in $d/2$ steps. At the end of step $t \in [d/2]$, we
shall have a pairing $(P_1,\ldots , P_t)$ of the arc $[L_t, R_t]$. The size of $[L_t
, R_t]$ is always $2t$. The first pairing
contains only $P_1 = \{L_1, R_1\}$ with $L_1 = 1$ and $R_1 = 2$. Given $(P_1, \ldots, P_t)$ and $[L_t, R_t]$, define the
random pair $P_t+1$ (independently of previous choices) by 

$$
P_{t+1} = \begin{cases}
    \{L_t - 2, L_t - 1\} &\text{ with probability $1/3$}\\
    \{L_t - 1, R_t + 1\} &\text{ with probability $1/3$}\\
    \{R_t + 1, R_t + 2\} &\text{ with probability $1/3$}\\
\end{cases}
$$

Define
$$
[L_{t+1},R_{t+1}] = [L_t,R_t] \cup P_{t+1}.
$$

So, $L_{t+1}$ is either $L_t - 2$, $L_t - 1$ or $L_t$, each value is obtained with probability $1/3$, and similarly
(but not independently) for $R_{t+1}$.

The final pairing is
$P = (P_1, P_2, \ldots , P_{d/2}
)$.
Denote by $P \sim \calD_P$ a pairing distributed according to $\calD_P$.

Once a pairing $P$ has been obtained, a word $w\in \{-k,k\}^d$ is obtained by simply randomly assigning $+k$ and $-k$ to the indices of any pair $P_i$. More formally, for every
$t \in [d/2]$, if $P_t = \{i_t
, j_t\}$, let with probability $1/2$, independently of all other choices,
$$
w_{i_t} = +k \text{ and } w_{j_t} = -k,
$$
and with probability 1/2,
$$
w_{i_t} = -k \text{ and } w_{j_t} = +k.
$$
Denote by $w\sim \calD$ a word in $\{-1,1\}^n$ that is sampled using this procedure. We call such a word an \emph{arc-partition}. For a pair $P_t = \{i_t,j_t\}$, we refer to $i_t$ and $j_t$ as \emph{partners}. 

\begin{definition}[Arc-full-rank]\label{def:arc-full-rk}
We say that a polynomial $f$ that is set-multilinear over $X = (X_1, \ldots, X_{d})$ is \textbf{arc-full-rank} if for every arc-partition $w\in \{-k,k\}^d$, $\rk_w(f) = 1$.
\end{definition}

\subsubsection{Construction of an Arc-full-rank Polynomial}\label{subsubsec:arc-full-rank-poly}

Below, we describe a simple construction of a polynomial sized ABP that computes an arc-full-rank set-multilinear polynomial.
The high-level idea is to construct an ABP in which every path between start-node and end-node
corresponds to a specific execution of the random process which samples arc-partitions. Each node in the ABP corresponds to an arc $[L, R]$, which sends an edge to each of the nodes $[L - 2, R], [L -
1, R + 1]$ and $[L, R + 2]$. The edges have specially chosen labels that help guarantee full rank with respect to
every arc-partition. For simplicity of presentation, we allow the edges of the program to be labeled
by degree four set-multilinear polynomial polynomials over the corresponding subset of the variable partition. This assumption can be easily removed by replacing
each edge with a polynomial-sized ABP computing the corresponding degree four polynomial.

Formally, the nodes of the program are even-size arcs in the $d$-cycle, $d$ an even integer. The
start-node of the program is the empty arc $\emptyset$ and the end-node is the whole cycle $[d]$ (both are
“special” arcs). Let $X = (X_1,\ldots, X_{d})$ be a collection of sets of variables where each $|X_i| = n$, and similarly, let $Y = (Y_1,\ldots, Y_{d})$ be a distinct collection of sets of variables where each $|Y_i| = n$ (we shall refer to the $Y$-variables as \emph{auxiliary} variables). For $i$ and $j$ in $\{1,\ldots,d\}$, let $X_i \cdot X_j$ denote the inner-product quadratic form $\sum_{k = 1}^n x_{ik}x_{jk}$. Here, we shall assume that $X_i = \{x_{i,1},\ldots, x_{i,n}\}$ and $Y_i = \{y_{i,1},\ldots, y_{i,n}\}$.

Construct the branching program by connecting a node/arc of size $2t$ to three nodes/arcs of
size $2t + 2$. For $t = 1$, there is just one node $[1, 2]$, and the edge from start-node to it is labeled
$y_{1,2}y_{2,1}(X_0\cdot X_1)$. For $t > 1$, the node $[L, R] \supset [1, 2]$ of size $2t < d$ is connected to the three nodes:
$[L-2, R] , [L-1, R+ 1]$, and $[L, R+ 2]$. (It may be the case that the three nodes are the end-node.)
The edge labeling is: 
\begin{itemize}
    \item The edge  between $[L, R]$ and $[L-2, R]$ is labeled $y_{L-2,L-1}y_{L-1,L-2}(X_{L-2}\cdot X_{L-1})$.
    \item The
edge between $[L, R]$ and $[L - 1, R + 1]$ is labeled $y_{L-1,R+1}y_{R+1,L-1}(X_{L-1}\cdot X_{R+1})$.
    \item The edge between $[L, R]$
and $[L, R + 2]$ is labeled $y_{R+1,R+2}y_{R+2,R+1}(X_{R+1}\cdot X_{R+2})$.
\end{itemize}
  
Consider the ABP thus described, and the polynomial $G_{n,d}$ it computes. For every path $\gamma$
from start-node to end-node in the ABP, the list of edges along $\gamma$ yields a pairing $P$; every edge $e$
in $\gamma$ corresponds to a pair $P_e = \{i_e, j_e\}$ of nodes in $d$-cycle. Thus,

\begin{equation}\label{eq:poly-def}
  G_{n,d} = \sum_{\gamma} \prod_{e = \{i_e, j_e\} \in \gamma}y_{i_e,j_e}y_{j_e,i_e} \cdot (X_{i_e}\cdot X_{j_e}). 
\end{equation}

where the sum is over all paths $\gamma$ from start-node to end-node. 

\begin{remark}\label{rem:parings-paths}
There is in fact a one-to-one
correspondence between pairings $P$ and such paths $\gamma$ (this follows by induction on $t$). {Note that this is true only because pairings are tuples i.e., they are \emph{ordered} by definition. Otherwise, it is of course still possible to obtain the same \emph{set} of pairs in a given pairing using multiple different orderings.} The sum
defining $G_{n,d}$ can be thought of, therefore, as over pairings $P$.  
\end{remark}

The following statement summarizes the main useful property of $G_{n,d}$.

\begin{lem}[\cite{KushS23}]\label{lem:Fn-VBP-props}
Over any field $\F$ of characteristic zero, the polynomial $ G_{n,d}$ defined above is arc-full-rank as a set-multilinear polynomial in the variables $X$ over the field $\F(Y)$ of rational functions
in $Y$.
\end{lem}

\begin{proof}
    Let $w\sim \calD$ be an arc-partition. 
    We want to show that $\calM_w(G_{n,d})$ has full rank. 
    The arc-partition $w$ is defined from a pairing $P = (P_1,\ldots,P_{d/2})$ (though as discussed in \autoref{rem:parings-paths}, there could be multiple such $P$). 
    The pairing $P$ corresponds to a path $\gamma$ from start-node to end-node. 
    Consider the polynomial $f$ that is obtained by setting every $y_{i,j} = y_{j,i} = 0$ in $F$ such that $\{i,j\}$ is not a pair in $P$, and setting every $y_{i,j} = y_{j,i} = 1$ for every pair $\{i,j\}$ in $P$. 
    Then, it is easy to see that the only terms that survive in \autoref{eq:poly-def} correspond to paths (and in turn, pairings) which have the same underlying \emph{set} of pairs as $P$.
    As a consequence, $f$ is simply some non-zero constant times a polynomial which is full-rank (recall \autoref{obs:ip-full-rank}).
    $M_w(f)$ being full rank then implies that $M_w(G_{n,d})$ is also full-rank.
\end{proof}

\section{Separation between $\VP$ and $\sbp$}

In the theorem below, $F_{n,d}$ refers to the polynomial defined in Section \ref{sec:VP}. 
In this section, we prove \autoref{thm:large-deg-VP} and \autoref{thm:small-deg-VP} by first proving the following statement. 

\begin{lem}\label{lem:VP}
    Given large enough integers $d\leq n$, any $\sbp$ of max-width $s$ and support size $t$ computing $F_{n,d}(X_1,\ldots,X_d,Y_1\ldots,Y_d)$ must satisfy at least one of the following:
    \begin{itemize}
        \item either $t>e^{d/96}$,
        \item or $t\cdot s^q \geq n^{\frac{\sqrt{dq}}{20}}$, for any integer $q$ in the range $ [12\ln(2t\sqrt{d}), d/4]$.
    \end{itemize}  
\end{lem}

\begin{proof}
    Suppose that $t\leq e^{d/96}$ (so that the range in the theorem statement is indeed well-defined). 
    
    First, we observe that for any $\sbp$ computing $F_{n,d}(X,Y)$, we can view each summand as an ordered set-multilinear branching program with respect to only the $X$ variables. 
    In other words, by appropriately collapsing the layers labelled using the $Y$ variables, each summand is a set-multilinear branching program over the field $\F(Y)$ \emph{ordered} with respect to $(X_{\sigma(1)},\ldots, X_{\sigma(d)})$ for some permutation $\sigma\in S_d$. 
    It is easy to see that this collapsing process does not increase the width or the size of any summand in the branching program. 
    The edge labels do get altered however: the coefficients of the $X$ variables in any edge label can now be polynomials in the $Y$ variables (and therefore, field constants in $\F(Y)$).

    Let $A$ be such a set-multilinear branching program of width $s$ and depth $d$ that is \emph{ordered} with respect to $(X_1,\ldots, X_d)$\footnote{In general, $A$ may be ordered with respect to an arbitrary permutation $\sigma$, but the assumption that $\sigma$ is the identity permutation in the discussion that follows is without loss of generality.}. 
    Recall that this means that for each $\ell\in[d]$, all edges of the ABP from layer $\ell-1$ to layer $\ell$ are labeled using a linear form {in} $\mathit{X_\ell}$. 
Given a node $u$ in layer $i$ and a node $v$ in layer $j>i$ of $A$, define $g_{u,v}$ to be the polynomial computed by the ABP restricted to the layers $i+1,\ldots,j-1$ with the source and the sink defined by $u$ and $v$ respectively. Consider an integer $q$ in the range specified in the theorem statement and let $r$ be the largest  integer $r$ such that the product $qr< d$ i.e., we must have $d-qr\leq q$. Consider the following decomposition of $A$:
\[
A(X_1,\ldots,X_d) = \sum_{u_1,\ldots, u_{q}} \prod_{i = 1}^{q+1} g_{u_{i-1},u_i},
\]
where $u_0$ and $u_{q+1}$ are defined to be the source and the sink of $A$ respectively, and for $1\leq i \leq q$, $u_i$ varies over all choices of nodes in layer $r\cdot i$. Note that hence, this expression contains at most $s^q$ terms. Also, note that each $g_{u_{i-1},u_i}$ is set-multilinear over the partition $(X_i)_{i\in S_i}$ where $S_i$ is the set $\{r(i-1) + 1, \ldots, ri\}$\footnote{If $A$ is instead ordered with respect to $\sigma$, then $S_i$ is taken to be the set $\{r(i-1) + 1, \ldots, ri\}$.} of length exactly $r$ for $i\in [q]$, and $S_{q+1}=\{rq+1,\ldots,d\}$ has length at most $q$. We now analyze the relative rank of each summand. 

Let  $w\in \{-k,k\}^d$ (where $k = \log n$) be an arbitrary word. By \autoref{clm:rk-props}, we see that 

\[
\rk_w\left(\prod_{i = 1}^{q+1} g_{u_{i-1},u_i}\right) = \prod_{i = 1}^{q+1} \rk_{w|_{S_i}}(g_{u_{i-1},u_i})
\leq \prod_{i = 1}^{q} \rk_{w|_{S_i}}(g_{u_{i-1},u_i})\leq \prod_{i = 1}^q 2^{-|w_{S_i}|/2}= 2^{-\frac12\left(\sum_{i=1}^q |w_{S_i}|\right)},
\]
from which we observe that the task of upper bounding this rank can be reduced to the task of lower bounding the sum $\sum_{i=1}^{q}|w_{S_i}|$, which is established below. 

Choose $w$ from $\{-k,k\}^d$ (where $k = \log n$) uniformly at random. For each $i\in[q]$, let $E_i$ denote the (bad) event that $|w_{S_i}|\leq  \sqrt{r}k/4$. Since $S_i$ is an interval of length $r$, by a standard estimation of binomial coefficients,  
we obtain that $\Pr[E_i]\leq 1/4$. Then,  by the Chernoff bound (\autoref{thm:Chernoff}), the probability that at least half of the events $E_i$ occur is at most $e^{-\frac{q}{12}}$. Therefore, with probability at least $1 - e^{-\frac{q}{12}}$, 
\[
\sum_{i=1}^q |w_{S_i}| \geq {q\sqrt{r}k/8},
\]
and therefore, by the sub-additivity of $\rk_w(\cdot)$,
\[
\rk_w(A) \leq s^q 2^{-\frac{kq\sqrt{r}}{16}} 
= s^q n^{-\frac{{q\sqrt{r}}}{16}}.
\]

Now, let $\sum_{i=1}^t A_i$ be a $\sbp$ computing $F_{n,d}$ with max-width bounded by $s$, and such that each $A_i$ is ordered set-multilinear with respect to the variable partition $(X_{\sigma_i(1)},\ldots, X_{\sigma_i(d)})$ for some permutation $\sigma_i\in S_d$. 
By the union bound and the discussion above, it follows that with probability at least $1-t\cdot e^{-\frac{q}{12}}$, 
\[
\rk_w(F_{n,d}) = \rk_w\left(\sum_{i=1}^t A_i\right) \leq t\cdot s^q n^{-\frac{{q\sqrt{r}}}{16}}.
\]

But now, we can condition on the event that $w_{[d]} = 0$ (which occurs with probability $\Theta(\frac{1}{\sqrt{d}}$)) to establish the existence of a word $w\in \{-k,k\}^d$ with $w_{[d]} = 0$ such that $w$ satisfies $
\rk_w(P)\leq t\cdot s^q n^{-q\sqrt{r}}$. This is because of the given bound $q\geq 12\ln(2t\sqrt{d})$. Because $\rk_w(F_{n,d}) = 1$ for such a $w$ by \autoref{lem:Fn-VP-props}, we conclude that $t\cdot s^q \geq n^{\frac{{q\sqrt{r}}}{16}} \geq n^{\frac{\sqrt{(d-q)q}}{16}}\geq n^{\frac{\sqrt{dq}}{20}}$, where the last inequality follows from our choice of $q$.
    
\end{proof}

\begin{proof}[Proof of \autoref{thm:large-deg-VP}]
We invoke Lemma \ref{lem:VP} with $d = n$. If $t \geq 2^{n^{1/3}}$, then we trivially have that the total-width is at least  ${\exp(\Omega(n^{1/3}))}$, so assume $t \leq 2^{n^{1/3}}$. We shall show that then, $s = \exp(\Omega(n^{1/3}))$, which will yield the desired result.

Set $q = \lceil 15n^{1/3}\rceil$. Then clearly, $q\leq n/4$. Moreover, as $t \leq 2^{n^{1/3}}$ by assumption, we verify that 
\[
q = \lceil 15n^{1/3} \rceil> 12 (n^{1/3} + \ln(2\sqrt{d})) \geq 12\ln(2t\sqrt{n}).
\]

Therefore, we can use \autoref{lem:VP} to obtain the inequality $t\cdot s^q \geq n^{\frac{\sqrt{nq}}{20}}$ (as $t\leq 2^{n^{1/3}}< e^{n/96}$ for large enough $n$). Plugging in $q = \lceil 15n^{1/3} \rceil$, we see that 
\[
s \geq \frac{n^{\frac{\sqrt{n/q}}{20}}}{t^{1/q}} \geq \frac{n^{\frac{n^{1/3}}{80}}}{2^{1/15}} = \exp(\Omega(n^{1/3})).
\]
\end{proof}

\begin{proof}[Proof of \autoref{thm:small-deg-VP}]
We consider cases as follows:

\paragraph{\underline{Case $t = \poly(n)$:}} Suppose there is a constant $c$ such that $t\leq n^c$. Set $q = 20 \ln (2n^c\sqrt{d})$ and note that $q = \Theta(\log n)$. We see that by \autoref{lem:VP},
\[
s \geq \frac{n^{\frac{\sqrt{d/q}}{20}}}{t^{1/q}} \geq n^{\frac{1}{20}\sqrt{\frac{d}{q}} - \frac{c}{q}}.
\]
Note that $c/q <1$ and decays to zero as $d$ becomes larger. Furthermore, as $d = \omega(\log n)$ by assumption and $q = \Theta(\log n)$, $d/q = \omega(1)$. We conclude that if $t$ is bounded by a polynomial in $n$, then $s$ must be super-polynomial in $n$.

\paragraph{\underline{Case $s = \poly(n)$:}} Suppose there is a fixed constant $c\geq 1$ such that $s\leq n^c$. Assume that $t<e^{d/96c^2}$, and set $q = d/1600c^2$. Then $q$ indeed lies in the range to apply \autoref{lem:VP}. We obtain the inequality
\[
t\geq \left(\frac{n^{\frac{\sqrt{d/q}}{20}}}{s}\right)^q \geq n^{\frac{\sqrt{dq}}{20} - cq} = n^{cq} = n^{\Omega(d)},
\]
which contradicts the assumption that $t<e^{d/96c^2}$. Hence, $t \geq e^{d/96c^2} = 2^{\Omega(d)}$ which is indeed super-polynomial in $n$ whenever $d = \omega(\log n)$.

Thus, in either case, it is shown that both $s$ and $t$ cannot be polynomially bounded. Hence, the total-width of any $\sbp$ computing $F_{n,d}$ cannot be polynomially bounded.
\end{proof}

\section{Tightness of ABP Set-Multilinearization}\label{sec:tightness}

In this section, we prove \autoref{thm:tightness} and in the process, also prove \autoref{thm:large-deg-VBP} and \autoref{thm:small-deg-VBP}. We first establish the following technical lemma that will be essential for these proofs. 

\begin{lem}\label{lem:VBP}
Let $d\leq n$ be growing parameters satisfying $d = \omega(\log n)$. There exist fixed constants $\gamma,c,c_1,c_2>0$ and $\eps\geq 1/100$ such that any $\sbp$ of max-width $s$ and support size $t$ computing $G_{n,d}(X_1,\ldots,X_d,Y_1\ldots,Y_d)$ must satisfy at least one of the following:
\begin{itemize}
    \item either $t\geq 2^{cc_2d}$,
    \item or $t\cdot s^q \geq n^{{{\gamma q(d/q)^\eps}}}$, for any integer $q$ in the range $ [\max\{c_1,(\log t)/c\}, c_2d]$.
\end{itemize}
\end{lem}

\begin{proof}
First, similar to the proof of \autoref{lem:VP}, we observe that for any $\sbp$ computing $G_{n,d}(X,Y)$, we can view each summand as an ordered set-multilinear branching program with respect to only the $X$ variables. In other words, by appropriately collapsing the layers labelled using the $Y$ variables, each summand is a set-multilinear branching program over the field $\F(Y)$ \emph{ordered} with respect to $(X_{\sigma(1)},\ldots, X_{\sigma(d)})$ for some permutation $\sigma\in S_d$. It is easy to see that this collapsing process does not  increase the width or the size of any summand branching program.  The edge labels do get altered however: the coefficients of the $X$ variables in any edge label can now be polynomials in the $Y$ variables (and therefore, field constants in $\F(Y)$).

Let $A$ be such a set-multilinear branching program of width $s$ and depth $d$ that is \emph{ordered} with respect to $(X_1,\ldots, X_d)$\footnote{In general, $A$ may be ordered with respect to an arbitrary permutation $\sigma$, but the assumption that $\sigma$ is the identity permutation in the discussion that follows is without loss of generality.}. Recall that this means that for each $\ell\in[d]$, all edges of the ABP from layer $\ell-1$ to layer $\ell$ are labeled using a linear form {in} $\mathit{X_\ell}$. Given a node $u$ in layer $i$ and a node $v$ in layer $j>i$ of $A$, define $g_{u,v}$ to be the polynomial computed by the ABP restricted to the layers $i+1,\ldots,j-1$ with the source and the sink defined by $u$ and $v$ respectively. Consider an integer $q$ in the range specified in the lemma statement and let $r$ be such that\footnote{\label{foot:rdq} If $q$ does not divide $d$, then we can let $r$ be $\lfloor d/q\rfloor$. In the discussion that immediately follows, we simply bound the relative rank of the `last' component by $1$ and so, the remaining analysis is nearly identical.} $qr = d$. Consider the following decomposition of $A$:
\[
A(X_1,\ldots,X_d) = \sum_{u_1,\ldots, u_{q-1}} \prod_{i = 1}^{q} g_{u_{i-1},u_i},
\]
where $u_0$ and $u_{q}$ are defined to be the source and the sink of $A$ respectively, and for $1\leq i \leq q-1$, $u_i$ varies over all choices of nodes in layer $r\cdot i$. Note that hence, this expression contains at most $s^{q-1}\leq s^q$ terms. Also, note that each $g_{u_{i-1},u_i}$ is set-multilinear over the partition $(X_i)_{i\in S_i}$ where $S_i$ is the set $\{r(i-1) + 1, \ldots, ri\}$\footnote{If $A$ is instead  ordered with respect to $\sigma$, then $S_i$ would be simply defined as the set $\{\sigma(r(i-1) + 1), \ldots, \sigma(ri)\}$.} of length exactly $r$. We now analyze the relative rank of each summand. 

By \autoref{clm:rk-props}, we see that for every appropriate word $w$,
\[
\rk_w\left(\prod_{i = 1}^{q+1} g_{u_{i-1},u_i}\right) = \prod_{i = 1}^{q+1} \rk_{w|_{S_i}}(g_{u_{i-1},u_i})
\leq \prod_{i = 1}^{q} \rk_{w|_{S_i}}(g_{u_{i-1},u_i})\leq\prod_{i = 1}^q  2^{-|w_{S_i}|/2}\leq 2^{-\frac12\left(\sum_{i=1}^q |w_{S_i}|\right)},
\]
from which we observe that the task of upper bounding this rank can be reduced to the task of lower bounding the sum $\sum_{i=1}^{q}|w_{S_i}|$, which is established below. 

Choose $w$ from the distribution $\calD$, as described in Section \ref{subsec:VBP}. We now view the partition $(S_1,\ldots,S_q)$ of $[d]$ as a fixed ``coloring'' of the latter set (and in turn, the $d$-cycle, as described in Section \ref{subsec:VBP}) i.e., each node $i\in[d]$ is assigned the color $k$ if and only if $i\in S_k$. For a pairing $P$ and set $S_k$, define the number of $k$-violations by
$$
V_k(P) = \{P_t \in P : |P_t \cap S_k| = 1\}.
$$
In words, it is the set of pairs in which one color is $k$ and the other color is different from $k$. For some fixed $0<\eps\leq 1/100$, denote
$$
\calG(P) = \{k \in [q] : |V_k(P)| \geq r^{2\eps}\}.
$$

Next, we state a technical lemma that states that with probability exponentially close to $1$, ``many" colors have ``many" violations. The constants $c_1,c_2$ that appear in the statement below indeed define the constants $c_1,c_2$ that are mentioned in the statement of Lemma \ref{lem:VBP}.
\begin{lem}[Many Violations Lemma]\label{lem:many-viol} Let $d\leq n$ be growing parameters satisfying $d = \omega(\log n)$. There exist fixed constants 
$0<\alpha,\beta<1$ and $c_1>0$, $0<c_2\leq 1$ such that for all integers $q$ in the range $ [c_1,c_2d] $ the following holds:
Let $S = (S_1, \ldots, S_q)$ be a partition of the $d$-cycle where each $|S_i| = r$.
Then,
$$
\Pr[|\calG(P)| \leq \alpha q] \leq r^{-\beta q},
$$
where $P\sim \calD_P$.
\end{lem}

We now show, in the claim below, how the preceding lemma can be used to argue that with probability exponentially close to $1$, an arc-partition $w$ exhibits large bias. The constant $c$ that appears below defines the constant $c$ in the statement of Lemma \ref{lem:VBP}.

\begin{claim}\label{clm:high-bias-VBP}
There exists a fixed constant $c>0$ such that
\[   \Pr\left[\sum_{i=1}^q |w_{S_i}| \leq \frac{\alpha qr^\eps k}{64}\right] \leq 2^{-cq},
\]
where the probability is over the choice of $w\sim \calD$.
\end{claim}
\begin{proof}
Let $\calE$ denote the event $\sum_{i=1}^q |w_{S_i}| \leq \frac{\alpha qr^\eps k}{64}$, and $\calA$ denote the event $|\calG(P)| > \alpha q$. From the law of total probability, it follows that $\Pr[\calE]\leq \Pr[\calE|\calA] + \Pr[\overline{\calA}]$, where $\overline{\calA}$ denotes the complement of $\calA$ -- therefore, it suffices to bound $\Pr[\calE|\calA]$. 

Fix a pairing $P\sim \calD_P$ such that the high probability event $\calA$ occurs. Consider an ordering $\sigma$ of the colors in $\calG(P)$. A color $\ell$ is said to be \emph{bright} with respect to an ordering if there are at least $r^{2\eps}/2$ nodes $x$ of color $\ell$ such that the partner of $x$ is colored using a color that appears \emph{after} $\ell$ in the ordering $\sigma$.  Call an ordering $\sigma$ of the nodes in $\calG(P)$ \emph{good} if there are at least $|\calG(P)|/2$ bright colors  with respect to $\sigma$. The observation is that for any ordering $\sigma$ of the colors, either $\sigma$ itself is good, or its reverse is good. We conclude that given any pairing $P$, there exists a good ordering of $\calG(P)$. Fix any such good ordering and let $\calH(P)$ be the collection of bright colors with respect to this ordering. Let the colors in $\calH(P)$ according to this good ordering be $\ell_1,\ldots, \ell_{q'}$.

Next, notice that if the sum $\sum_{j=1}^{q}|w_{S_j}|$ is at most $\frac{\alpha qr^\eps k}{64}$, then so is the sum $\sum_{\ell\in \calH(P)}|w_{S_\ell}|$. Let $q' = |\calH(P)|$ (which is at least $\alpha q/2$ if $|\calG(P)|>\alpha q$). View the sampling of $w$ from $P$ as happening in a specific order, according to the order
of $\ell_1, \ell_2, \ldots, \ell_{q'}$: First define $\Pi$ on pairs with at least one point with color $\ell_1$, then define $\Pi$ on
remaining pairs with at least one point with color $\ell_2$, and so forth. When finished with $\ell_1,\ldots, \ell_{q'}$,
continue to define $\Pi$ on all other pairs.

Observe that if the sum $\sum_{\ell\in \calH(P)}|w_{S_\ell}|$ is at most $\frac{\alpha qr^\eps k}{64}$, then there exists a subset $T\subseteq \calH(P)$ with half its size, $|T|= |\calH(P)|/2$,\footnote{We assume without loss of generality that $|\calH(P)|$ is even to avoid ceilings and floors.} such that every color $\ell$ in $T$ satisfies $|w_{S_\ell}|\leq kr^\eps /16$ (otherwise, we obtain an immediate contradiction). There are at most $2^{|\calH(P)|}$ many choices for such $T$ -- fix such a choice $T$ and relabel the colors in $T$ as $\ell_1,\ldots, \ell_{q''}$, where $q'' = |\calH(P)|/2$ and this order respects the order described in the paragraph above.
For every $\ell_j\in T$, define $E_j$ to be the event that $|w_{S_{\ell_j}}|\leq kr^\eps /16$. 
By choice, conditioned on $E_1,\ldots, E_{j-1}$, there are at least $r^{2\eps}/2$ pairs $P_t$ so that
$|P_t \cap S_{\ell_j}| = 1$ that are not yet assigned a `positive' or `negative' sign (with a magnitude equal to $|k|$). For every such $P_t$, the element in $P_t \cap S_{\ell_j}$
is assigned a positive sign with probability $1/2$, and is independent of any other $P_{t'}$. Therefore, $\Pr[E_j]$ is bounded by the probability that a binomial random variable over a universe of size $r^{2\eps}/2$ lies in any \emph{specific} interval of size $r^\eps/8$. By a standard estimation of binomial coefficients, this probability is bounded by a constant, $1/5$.

Hence, for any fixed choice of $T\subseteq \calH(P)$, for all $\ell_j \in T$,
\[
\Pr[E_j|E_1,\ldots,E_{j-1},P] \leq 1/5.
\]
Therefore, for any fixed choice of $T\subseteq \calH(P)$, by the chain rule, it follows that 
\[
\Pr[\cap_{j\in T}E_j]\leq 5^{-|T|} = 2^{-(\log 5)|\calH(P)|/2}.
\]
There are at most $2^{|\calH(P)|}$ choices for $T$. We conclude that 
\[\Pr[\calE|\calA]\leq 2^{|\calH(P)|} \cdot 2^{-(\log 5)|\calH(P)|/2}\leq 2^{-|\calH(P)|/7}\leq 2^{-\alpha q/14}.
\]
Finally, we note that by the many violations lemma (\autoref{lem:many-viol}), $\Pr[\overline{\calA}]\leq r^{-\beta q}\leq 2^{-\beta q}$. Thus, $\Pr[\calE] \leq \Pr[\calE|\calA] + \Pr[\overline{\calA}] \leq 2^{-\alpha q/14} + 2^{-\beta q} \leq 2^{-cq}$ for some fixed constant $c>0$ which can be defined in terms of $\alpha$ and $\beta$.
\end{proof}

From the claim, it follows that with probability at least $1 - 2^{-cq}$, 
\[
\sum_{i=1}^q |w_{S_i}| \geq \frac{\alpha qr^\eps k}{64},
\]
and therefore, by the sub-additivity of $\rk_w(\cdot)$,
\[
\rk_w(A) \leq s^q 2^{-\frac{\alpha kqr^\eps }{128}} = s^q n^{-{\frac{\alpha qr^\eps }{128}}}.
\]

Now, let $\sum_{i=1}^t A_i$ be a $\sbp$ computing $G_{n,d}$ with max-width bounded by $s$, and such that each $A_i$ is ordered set-multilinear with respect to the variable partition $(X_{\sigma_i(1)},\ldots, X_{\sigma_i(d)})$ for some permutation $\sigma_i\in S_d$. Assume that $2^{cc_1}< t< 2^{cc_2d}$ (if the second inequality does not hold, then the first item in the lemma statement is true; and we deal with the case $t\leq 2^{cc_1}$ at the end of the proof). 
By the union bound and the discussion above, it follows that with probability at least $1-t\cdot 2^{-cq}$, 
\[
\rk_w(F_{n,d}) = \rk_w\left(\sum_{i=1}^t A_i\right) \leq t\cdot s^q n^{-{\frac{\alpha qr^\eps }{128}}}.
\]
Since $q> (\log t)/c$ by assumption, note that $1-t\cdot 2^{-cq}>0$. Furthermore, since we are assuming $2^{cc_1}< t$, we have that $q\geq c_1$ and therefore, Lemma \ref{lem:many-viol} applies and so does \autoref{clm:high-bias-VBP} along with the entire discussion above. We conclude that there exists an arc-partition $w\in \{-k,k\}^d$ such that $w$ satisfies $
\rk_w(F_{n,d})\leq t\cdot s^q n^{-{\frac{\alpha qr^\eps }{128}}}$. Because $\rk_w(F_{n,d}) = 1$ for such a $w$ by \autoref{lem:Fn-VBP-props}, we conclude that $t\cdot s^q \geq n^{{\frac{\alpha qr^\eps }{128}}}$.

Finally, if $t$ is bounded by the constant $2^{cc_1}$, then we just add constantly many width-$2$ ordered set-multilinear branching programs which each compute the zero polynomial so that the new support size of the sum becomes larger than $2^{cc_1}$ and the previous case applies, and we are able to conclude $(t + 2^{cc_1})\cdot s^q \geq n^{{\frac{\alpha qr^\eps }{128}}}$. Appropriately defining the constant $\gamma>0$ then lets  us conclude the desired bound $t\cdot s^q \geq n^{{{\gamma q(d/q)^\eps}}}$.
\end{proof}

\begin{proof}[Proof of \autoref{thm:large-deg-VBP}]
Set $\delta = \eps/(1+\eps)$, where $\eps$ is as defined in the Lemma \ref{lem:VBP}. If $t \geq 2^{n^{\delta}}$, then we trivially have that the total-width is at least  ${\exp(\Omega(n^{\delta}))}$, so assume $t \leq 2^{n^{\delta}}$ (so that the second item of \autoref{lem:VBP} applies). We shall show that then, $s = \exp(\Omega(n^{\delta}))$, which will yield the desired result.

By \autoref{lem:VBP}, any $\sbp$ of max-width $s$ and support size $t$ computing $G_{n,n}$ satisfies the inequality $t\cdot s^q \geq n^{{\gamma q(n/q)^\eps}}$, for any integer $q$ in the range $ [(\log t)/c, c_2n]$. Set $q = \lceil n^{\delta}/c\rceil$. Then clearly, $q\geq (\log t)/c$ and so, $q$ lies in the required range.  Plugging in the setting for $q$ in the inequality, we see that 
\[
s \geq \frac{n^{\gamma{({n/q})^\eps}}}{t^{1/q}} \geq \frac{n^{\gamma({cn^{1-\delta}})^{\eps}}}{2^c} = \exp(\Omega(n^{\delta})).
\]
\end{proof}

\begin{proof}[Proof of \autoref{thm:small-deg-VBP}]
We consider cases as follows:
\paragraph{\underline{Case $t = \poly(n)$:}} Suppose there is a constant $c'$ such that $t\leq n^{c'}$. Set $q = \lceil(\log n^{c'})/c\rceil = \lceil (c'\log n)/c\rceil$ and note that $q = \Theta(\log n)$. We see that by \autoref{lem:VBP},
\[
s \geq \frac{n^{\gamma{({d/q})^\eps}}}{t^{1/q}}\geq n^{{\gamma (d/q)^\eps}- \frac{c'}{q}}.
\]
Note that $c'/q <1$ and decays to zero as $d$ becomes larger. Furthermore, as $d = \omega(\log n)$ by assumption and $q = \Theta(\log n)$, $d/q = \omega(1)$. We conclude that if $t$ is bounded by a polynomial in $n$, then $s$ must be super-polynomial in $n$.

\paragraph{\underline{Case $s = \poly(n)$:}} Suppose there is a fixed constant $c'\geq 1$ such that $s\leq n^{c'}$. Define $r$ to be the constant $ \max\{(\frac{2c'}{\gamma})^{1/\eps}, \frac{1}{c_2}\}$. Set $q = d/r$ and assume that $t<2^{cq}$. Then $q$ indeed lies in the range to apply \autoref{lem:VBP}: we have $q\leq c_2d$ because $1/r\leq c_2$ by definition, and $q\geq (\log t)/c$ because of the assumption on $t$. We obtain the inequality
\[
t\geq \left(\frac{n^{\gamma ({d/q})^\eps}}{s}\right)^q \geq \left(\frac{n^{{\gamma r^\eps}}}{n^{c'}}\right)^q\geq \left(\frac{n^{\frac{2c'\gamma }{\gamma}}}{n^{c'}}\right)^q = n^{c'q} = n^{\Omega(d)},
\]
which contradicts the assumption that $t<2^{cq}$. Hence, $t \geq 2^{cq} = 2^{\Omega(d)}$ which is indeed super-polynomial in $n$ whenever $d = \omega(\log n)$.

Thus, in either case, it is shown that both $s$ and $t$ cannot be polynomially bounded. Hence, the total-width of any $\sbp$ computing $G_{n,d}$ cannot be polynomially bounded.
\end{proof}

Finally, we observe that \autoref{thm:tightness} follows immediately from (i) the ABP construction of $G_{n,d}$ given in Section \ref{subsec:VBP}, and (ii), the proof of the $s = \poly(n)$ case in \autoref{thm:small-deg-VBP} above.

\section{Optimal Separation between Ordered and Unordered smABPs}\label{sec:opt-sep-sml-osml}
In this section, we prove \autoref{thm:sep-ord-unord}. The result is that $G_{n,d}(X_1,\ldots, X_d,Y_1,\ldots,Y_d)$, as described in Section \ref{subsec:VBP}, does not have small \emph{ordered} set-multilinear branching programs with respect to \emph{any} ordering of the $X_i$s and $Y_j$s. 
More precisely, we claim that any set-multilinear branching program computing $G_{n,d}$ which is ordered with respect to some permutation of the sets in the collection $\calS = \{X_1,\ldots, X_d,Y_1,\ldots,Y_d\}$ must have size at least $n^{\Omega(d)}$.

\begin{proof}[Proof of \autoref{thm:sep-ord-unord}]
    Let $A$ be a set-multilinear branching program computing $G_{n,d}$, which is ordered with respect to a permutation $\pi$ of the sets in $\calS$.
    That is $\pi: [2d]\rightarrow \calS$ is a bijective map where $\pi(i) = Z$, for some $Z\in \calS$, if all edges between layer $i-1$ and layer $i$ are labelled using linear forms in the variables of $Z$. 
    Let $\pi_X:[d]\rightarrow \{X_1,\ldots,X_d\}$ be the ordering inherited from $\pi$ by the $X$ sets, that is $\pi_X(i) = X_j$ if $X_j$ is the $i$-th $X$-set appearing in the sequence $\pi(1),\ldots,\pi(2d)$. 
    Define a word $w\in \{-k,k\}^d$ using $\pi_X$ as follows: let 
    \[
        w_i = k \text{ if and only if } \pi_X^{-1}(X_i)\in [d/2].
    \] 
    That is, the edges labelled using linear forms in $X_i$ appear on the `left' half in $A$, when the $Y$ sets are ignored. 
    Note that $w$ is \emph{symmetric}, or in other words, $w_{[d]} = 0$.

    Now, for $i\in[d]$, define $s_i(A)$ to be the size of layer $\pi^{-1}(\pi_X(i))$ in $A$.
    The technique used by Nisan \cite{Nisan91} can be used directly to show the following.

    \begin{lem}[\cite{Nisan91}]\label{lem:nisan-abpsize}
        If $G_{n,d}(X_1,\ldots,X_d)\in \mathbb{K}[X_1,\ldots,X_d]$ is thought of as a polynomial only over the $X$ variables (here $\mathbb{K} = \F(Y)$, the field of rational functions in $Y$), then $n^{d/2}\cdot \rk_w(G_{n,d})\leq s_{d/2}(A)$.\footnote{The $n^{d/2}$ factor comes from using relative-rank rather than rank.}
    \end{lem}

    Thus, in order to prove a $n^{\Omega(d)}$ lower bound on the width of $A$, it suffices to lower bound $\rk_w(G)$ by $n^{-\alpha d}$ for some $0 < \alpha < 1/2$.
    We prove such a lower bound using the following lemma.

    \begin{lem}\label{lem:overlap}
        Given any $w\in\{-k,k\}^d$ with $w_{[d]} = 0$, there exists an arc-partition $v\in\{-k,k\}^d$ obtained from a pairing $P = (P_1,\ldots,P_{d/2})$ such that $w$ `splits' a constant fraction of the pairs in $P$: more precisely, there is a set $S\subset P$ of size at least $d/8$ such that if $(i,j) \in S$, then $w_i + w_j = 0$.
    \end{lem}

    Let us first prove the lower bound on $\rk_w(G_{n,d})$ assuming this lemma. 
    Consider the polynomial $f$ that is obtained by setting every $y_{i,j} = y_{j,i} = 0$ in $G_{n,d}$ such that $\{i,j\}$ is not a pair in $P$, and setting every $y_{i,j} = y_{j,i} = 1$ for every pair $\{i,j\}$ in $P$ (where $P$ is as in \autoref{lem:overlap}). 
    Then, it is easy to see that the only terms that survive in \autoref{eq:poly-def} correspond to paths (and in turn, pairings) which have the same underlying \emph{set} of pairs as $P$. 
    As a consequence,
    \[
        f(X)= c_P \cdot \prod_{t = 1}^{d/2} (X_{i_t}\cdot X_{j_t})
    \]
    for a non-zero constant $c_P$.
    Next, notice that the rank of the matrix $M_w(f)$ (over the field $\F$) serves as a lower bound on the rank of the matrix $M_w(G_{n,d})$ (over the field $\F(Y)$).
    Indeed, if the former is $r$ then there is a non-vanishing $r\times r$ minor of $M_w(f)$. 
    But this implies that the determinant of the corresponding $r\times r$ sub-matrix in $M_w(G_{n,d})$ must be a non-zero polynomial in $\F[Y]$ as it has a non-zero evaluation.
    From this we conclude that $M_w(G_{n,d})$ also has a non-vanishing $r\times r$ minor and therefore has rank at least $r$. 
    Stated in terms of relative rank, we have showed that $\rk_w(f)\leq \rk_w(G_{n,d})$.

    Now, given the set $S\subset [d/2]$ from \autoref{lem:overlap}, we can lower bound the relative rank of $M_w(f)$ as follows: first, note that if for some $t\in[d/2]$, the pair $P_t$ of $P$ is in $S$, then by \autoref{obs:ip-full-rank}, $\rk_{w|_{P_t}}(X_{i_t}\cdot X_{j_t}) = 1$. 
    Secondly, if $w$ does \emph{not} split some pair $P_t$ of $P$ (i.e., $w_{i_t} = w_{j_t}$), then $M_{w|_{P_t}}(X_{i_t}\cdot X_{j_t})$ is simply a one-dimensional vector (either $1\times n^2$ or $n^2\times 1$) and we trivially have $\rk_{w|_{P_t}}(X_{i_t}\cdot X_{j_t}) = 1/n$. 
    Combining, and using the multiplicativity of $\rk_w(\cdot)$ (third item of \autoref{clm:rk-props}), 
    \begin{align*}
        \rk_w(f(X)) = \rk_w(f(X)/c_P) &= \prod_{t = 1}^{d/2} \rk_{w|_{P_t}}(X_{i_t}\cdot X_{j_t})\\
        &= \prod_{t\in S} \underbrace{\rk_{w|_{P_t}}(X_{i_t}\cdot X_{j_t})}_{= 1} \prod_{t\in[d/2]\setminus S}\underbrace{\rk_{w|_{P_t}}(X_{i_t}\cdot X_{j_t})}_{\geq 1/n}
        \geq \frac{1}{n^{d/2 - |S|}}.
    \end{align*}
    
    Therefore, we have $\rk_w(G_{n,d})\geq \rk_w(f) \geq 1/n^{d/2 - |S|}$ and hence, by \autoref{lem:nisan-abpsize}, the size of $A$ is at least $s_{d/2}(A)\geq n^{d/2}\cdot \rk_w(G_{n,d}) \geq n^{|S|} = n^{\Omega(d)}$.
\end{proof}

We now give the proof of \autoref{lem:overlap}.

\begin{proof}[Proof of \autoref{lem:overlap}]
    Let $L_+, L_-, R_+, R_- \in [d]$ be defined as follows.
    \begin{align*}
        R_+ &= \left\{i \in  \left[ 3, \frac{d}{2}+1 \right] \quad : \quad w_i > 0 \right\}
        \qquad \qquad \qquad
        R_- = \left\{i \in  \left[ 3, \frac{d}{2}+1 \right] \quad : \quad w_i < 0 \right\}\\
        L_+ &= \left\{i \in  \left[\frac{d}{2}+2, d \right] \quad : \quad w_i > 0 \right\}
        \qquad \qquad \qquad
        L_- = \left\{i \in  \left[\frac{d}{2}+2, d \right] \quad : \quad w_i < 0 \right\}
    \end{align*}
    Clearly, either $|R_+|\geq\frac{d}{4}$ or $|R_-|\geq\frac{d}{4}$.
    Without loss of generality, let us assume that $|R_+|\geq\frac{d}{4}$.
    
    Also, without loss of generality\footnote{If this is not the case, then $|L_-| -2 \leq |R_+| \leq |L_-| + 2$ and we redefine $R_+, L_-$ to be the largest possible subsets of the originally defined $R_+, L_-$, respectively, such that $|R_+| = |L_-|$.} we can assume that $w_1 + w_2 = 0$.
    Then $|R_+| = |L_-|$, and say $|R_+| = p$.
    Further, let 
    \[
        R_+ = \{i_1, \ldots, i_p\} \text{ with } i_1 < \cdots < i_p \quad \text{ and } \quad L_- = \{j_1, \ldots, j_p\} \text{ with } j_1 > \cdots > j_p.
    \]
    We can then define an initial pairing, $P^0 = \{(i_\ell, j_\ell)\}_{\ell \in [p]}$.
    Let us also define the set $S^0 = \emptyset$.
    The goal is to iteratively update $P^0$ such that, at the end, we have a pairing corresponding to an arc-partition. 
    We will also update $S^0$ at each step so that, at the end, each pair in $S$ are of opposite signs.
    
    Let $u_0=2, v_0=1$.
    Intuitively, $u_\ell, v_\ell$ are the right most and left-most points, respectively, of the partial arc-partition, $P^\ell$, defined till the $\ell$-th iteration.
    Given $u_{\ell-1}, v_{\ell-1}, P^{\ell-1}, S^{\ell -1}$ for any $\ell \in [p]$, we will define $u_\ell, v_\ell, P^\ell, S^\ell$ as follows.
    Note that the calculations are $\pmod d$ when $\ell=1$.
    \begin{enumerate}
        \item[\underline{\textbf{Case 1:}} ]  $(i_\ell - u_{\ell-1})$ and $(v_{\ell-1} - j_\ell)$ are both odd.
        \[
            P^\ell = P^{\ell-1} \cup \{(u_{\ell-1}+1, u_{\ell-1}+2), \ldots, (i_\ell -2, i_\ell -1)\} \cup \{(v_{\ell-1}-2, v_{\ell-1}-1), \ldots, (j_\ell +1, j_\ell +2)\}
        \]
        \[
            u_\ell = i_\ell \qquad \qquad v_\ell = j_\ell \qquad \text{ and } \qquad S^\ell = S^{\ell -1} \cup \{(i_\ell, j_\ell) \}
        \]
        \item[\underline{\textbf{Case 2:}} ] $(i_\ell - u_{\ell-1})$ is even and $(v_{\ell-1} - j_\ell)$ is odd.
        
        We first define
        \[
            Q^\ell = \begin{cases}
                (P^{\ell-1} \setminus \{(i_\ell, j_\ell)\}) & \cup \quad \{(u_{\ell-1}+1, u_{\ell-1}+2), \ldots, (i_\ell -1, i_\ell)\} \\
                & \cup \quad \{(v_{\ell-1}-2, v_{\ell-1}-1), \ldots, (j_\ell +1, j_\ell +2\}, 
                \qquad \text{if } (i_\ell - u_{\ell-1}) > 0\\
                (P^{\ell-1} \setminus \{(i_\ell, j_\ell)\}) &\cup \quad \{(v_{\ell-1}-2, v_{\ell-1}-1), \ldots, (j_\ell +1, j_\ell +2)\} 
                \qquad \text{ if } (i_\ell - u_{\ell-1}) = 0,
            \end{cases}
        \]
        
        and then define
        \[
            P^\ell = \begin{cases}
                Q^\ell \cup \{(j_\ell -1, j_\ell)\} \quad \text{ if } \quad j_\ell -1 \in L_+.\\
                Q^\ell \cup \{(i_\ell +1, j_\ell)\} \quad \text{ otherwise.} 
                \end{cases} 
        \]
        
        Also,
        \[
            u_\ell = \begin{cases}
                i_\ell \qquad \quad \text{ if } \quad j_\ell -1 \in L_+\\
                i_\ell +1 \quad \text{ otherwise.} 
            \end{cases} 
            \qquad \qquad 
            v_\ell = \begin{cases}
                j_\ell - 1 \quad \text{ if } \quad j_\ell -1 \in L_+\\
                j_\ell \qquad \quad \text{otherwise.} 
            \end{cases} 
        \]
        and
        \[
            S^\ell = \begin{cases}
                S^{\ell -1} \cup \{(i_\ell, j_{\ell -1})\} \qquad \text{ if } (i_\ell - u_{\ell-1}) = 0  \\
                S^{\ell -1} \cup \{(i_\ell -1, i_\ell)\} \qquad \text{otherwise.}
            \end{cases}
        \]
        \item[\underline{\textbf{Case 3:}}] $(i_\ell - u_{\ell-1})$ is odd and $(v_{\ell-1} - j_\ell)$ is even.

        We first define
        \[
            Q^\ell = \begin{cases}
                (P^{\ell-1} \setminus \{(i_\ell, j_\ell)\}) &\cup \quad \{(u_{\ell-1}+1, u_{\ell-1}+2), \ldots, (i_\ell -2, i_\ell -1)\} \\
                &\cup \quad \{(v_{\ell-1}-2, v_{\ell-1}-1), \ldots, (j_\ell, j_\ell +1)\} 
                \qquad \quad \text{ if } (v_{\ell-1} - j_\ell) > 0, \\
                (P^{\ell-1} \setminus \{(i_\ell, j_\ell)\}) &\cup \quad \{(u_{\ell-1}+1, u_{\ell-1}+2), \ldots, (i_\ell -2, i_\ell -1)\} \qquad \text{if } (v_{\ell-1} - j_\ell) = 0,
            \end{cases}
        \]
        
        and then define
        \[
            P^\ell = \begin{cases}
                Q^\ell \cup \{(i_\ell, i_\ell +1)\} \quad \text{ if } \quad i_\ell +1 \in R_-.\\
                Q^\ell \cup \{(i_\ell, j_\ell -1)\} \quad \text{ otherwise.} 
                \end{cases} 
        \]
        
        Also,
        \[
            u_\ell = \begin{cases}
                i_\ell +1 \quad \text{ if } \quad i_\ell +1 \in R_-\\
                i_\ell \qquad \quad \text{otherwise.} 
            \end{cases} 
            \qquad \qquad 
            v_\ell = \begin{cases}
                j_\ell \qquad \quad \text{if } \quad i_\ell +1 \in R_-\\
                j_\ell -1 \quad \text{ otherwise.} 
            \end{cases} 
        \]
        and
        \[
            S^\ell = \begin{cases}
                S^{\ell -1} \cup \{(i_{\ell -1}, j_\ell)\} \qquad \text{ if } (v_{\ell -1} - j_\ell) = 0.  \\
                S^{\ell -1} \cup \{(j_\ell, j_\ell +1)\} \qquad \text{otherwise.}
            \end{cases}
        \]

        \item[\underline{\textbf{Case 4:}}]  $(i_\ell - u_{\ell-1})$ and $(v_{\ell-1} - j_\ell)$ are both even.

        We first define
        \[
            Q^\ell = \begin{cases}
                (P^{\ell-1} \setminus \{(i_\ell, j_\ell)\}) \qquad \qquad \qquad \qquad  \qquad \qquad \qquad \qquad \text{ if } (i_\ell - u_{\ell-1}) = 0 \\
                (P^{\ell-1} \setminus \{(i_\ell, j_\ell)\}) \cup \{(u_{\ell-1}+1, u_{\ell-1}+2), \ldots, (i_\ell -1, i_\ell)\} \quad \text{ otherwise}
            \end{cases} 
        \]
        and then define
        \[
            P^\ell = \begin{cases}
                Q^\ell \qquad \qquad \qquad \qquad \qquad \qquad \qquad \qquad \text{ if } (v_{\ell-1} - j_\ell) = 0 \\
                Q^\ell \cup \{(v_{\ell-1}-2, v_{\ell-1}-1), \ldots, (j_\ell, j_\ell +1)\} \quad \text{ otherwise.}
            \end{cases}
        \]
        Also,
        \[
            u_\ell = i_\ell \qquad \qquad v_\ell = j_\ell \qquad \text{ and } \qquad S^\ell = \begin{cases}
                S^{\ell -1} \cup \{(i_\ell, j_{\ell -1})\} \qquad \text{ if } (i_\ell - u_{\ell-1}) = 0.  \\
                S^{\ell -1} \cup \{(i_\ell -1, i_\ell)\} \qquad \text{otherwise.}
            \end{cases}
        \]
    \end{enumerate}  
    Note that $(v_p - u_p)$ must be odd.
    So finally, we define $P = P^p \cup \{(u_p+1, u_p+2), \ldots, (v_p-2, v_p-1)\}$ and $S = S^p$.
    Firstly note that $P$ corresponds to an arc-partition, since $P^0$ is a partial arc-partition and it is easy to check that for each $\ell \in [p]$, $P^\ell$ is a valid extension of $P^{\ell-1}$.
    
    Since $p \geq d/8$, $S$ clearly has size at least $d/8$.
    The only thing left to complete the proof is to check that $S$ has the other required property.
    For any $\ell \in [p]$, let us assume that $S^{\ell -1} \subset P$ has the property that if $(i,j) \in S^{\ell -1}$, $w_i + w_j = 0$.    
    In Case 1, clearly, $S^\ell$ continues to have this property.
    In Cases 2 and 4, if $(i_\ell - u_{\ell-1}) > 0$, then again $S^\ell$ clearly continues to have this property.
    When $(i_\ell - u_{\ell-1}) = 0$, it must mean that $i_\ell = u_{\ell-1} = i_{\ell -1} + 1$ and so $i_{\ell -1} + 1 \not\in R_-$.
    That is, in the previous iteration Case 2 had been true and $(i_\ell, j_{\ell -1}) = (i_{\ell -1} + 1, j_{\ell -1})$ had been added to $P^{\ell -1}$.
    Therefore $S^\ell \subset P$ and, also, $w_{i_\ell} + w_{j_{\ell -1}} = 0$.
    A similar argument for Case 3 then completes the proof.
\end{proof}

\section*{Acknowledgements}
Parts of this work were done while the first author was visiting TIFR Mumbai and ICTS-TIFR Bengaluru, and she would like to thank Venkata Susmita Biswas, Ramprasad Saptharishi, Prahladh Harsha and Jaikumar Radhakrishnan for the hospitality. 
The first author would also like to thank Anamay Tengse for useful discussions while trying to understand \cite{RR20} and \cite{GR21}.

\bibliographystyle{alpha}
\bibliography{bibfile}

\appendix
\section{Proof of the Many Violations Lemma}\label{sec:app}
We briefly recall some notation from Section \ref{sec:tightness}. Recall that $w$ is chosen from the distribution $\calD$, as described in Section \ref{subsec:VBP}. For a pairing $P$, and a set $S_k$, we defined the number of $k$-violations as
$$
V_k(P) = \{P_t \in P : |P_t \cap S_k| = 1\}.
$$

In words, it is the set of pairs in which one color is $k$ and the other color is different. We used the following notation to denote the set of colors with ``many'' violations (for some fixed $0<\eps\leq 1/100$)
$$
\calG(P) = \{k \in [q] : |V_k(P)| \geq r^{2\eps}\}.
$$

As mentioned previously, this subsection is adapted from the proof of the (weaker) many violations lemma in \cite{KushS23}, which is in turn adapted from Lemma 4.1 in \cite{DvirMPY12}. 

\paragraph{Lemma.}(Many Violations Lemma restated) \textit{Let $d\leq n$ be growing parameters satisfying $d = \omega(\log n)$. There exist fixed constants 
$0<\alpha,\beta<1$ and $c_1>0$, $0<c_2\leq 1$ such that for all integers $q$ in the range $ [c_1,c_2d] $ the following holds:
Let $S = (S_1, \ldots, S_q)$ be a partition of the $d$-cycle where each $|S_i| = r$.\footnote{As explained in \autoref{foot:rdq} we assume w.l.o.g. $d=rq$.}
Then,
$$
\Pr[|\calG(P)| \leq \alpha q] \leq r^{-\beta q},
$$
where $P\sim \calD_P$.}

\begin{proof}
Set $\alpha = 1/1000$. Fix a partition (or a ``coloring'') $S = (S_1,\ldots, S_q)$ of the $d$-cycle satisfying the conditions of the lemma.
Think of $S$ as a function from the $d$-cycle to the set $[q]$, assigning every node its color in $[q]$. $S(i)$ is the
color of $i$. Use the following definition to partition the proof into cases. For a color $k$, count the
number of jumps in it (with respect to the partition $S$) to be
\[
J_k = \{j \in S_k : k = S(j) \neq S(j + 1)\},
\]
the set of elements $j$ of color $k$ so that $j + 1$ has a color different from $k$. 

\paragraph{Case 1: Many colors with many jumps.}  The high-level idea is that each color with many jumps
has many violations because pairs of the form $(j, j + 1)$ yield violations as soon as they are constructed.

Assume that for at least $q/2$ colors $k$,
$|J_k| > r^{4\eps}$.
Denote by $B \subseteq [q]$ the set of $k$’s that satisfy this inequality. Then, for every $k$ in $B$, there exists
a subset $Q_k \subset J_k$ of size $N = \lceil r^{4\eps}\rceil$. Let
\[
Q \coloneqq \bigcup_{k\in B} Q_k.
\]
We think of the construction of the (random) pairing $P$ as happening in \emph{epochs}, depending on $Q$, as
follows.

For $t > 0$, define the random variable
\[
Q(t) = Q \setminus [L_t - 4, R_t + 4],
\]
the set $Q$ after removing a four-neighborhood of $[L_t, R_t]$. 
For a certain sequence of time steps $t$, we will define special nodes $q_t$ which lie in this small `cloud' around the arc $[L_t,R_t]$ (i.e., within a distance of $4$ on either side of the arc) - it is for these special nodes $q_t$ that the set of pairs $(q_t,q_{t+1})$ will provide many violations. We now formalize this intuition. 

Let $\tau_1 \geq \tau_0 \coloneqq 1$ be the first time $t$ after $\tau_0$ so that the distance between $[L_t
, R_t
]$ and $Q(\tau_0)$ is
at most two. The distance between $[L_{\tau_0
}
, R_{\tau_0}
]$ and $Q(\tau_0)$ is at least five. The size of the arc $[L_t
, R_t
]$ increases by two at each time step. So, $\tau_1 \geq \tau_0+ 2$. Let $q_1$ be an element of $Q(\tau_0)$ that is of distance
at most two from $[L_{\tau_1}
, R_{\tau_1}
]$; if there is more than one such $q_1$, choose arbitrarily. The minimality
of $\tau_1$ implies that $q_1$ is not in $[L_{\tau_1}
, R_{\tau_1}
]$.

Let $\tau_2 \geq \tau_1$ be the first time $t$ after $\tau_1$ so that the distance between $[L_t
, R_t
]$ and $Q(\tau_1)$ is at most
two. Let $q_2$ be an element of $Q(\tau_1)$ that is of distance at most two from $[L_{\tau_2}
, R_{\tau_2}
]$.
Define $\tau_j$, $q_j$ for $j > 2$ similarly, until $Q(\tau_j )$ is empty. As long as $|Q(\tau_j )| \geq 8$, we have $|Q(\tau_j+1)| \geq
|Q(\tau_j )| - 8$. This process, therefore, has at least $qN/16$ steps.
For $1 \leq j \leq qN/16$, denote by $E_j$ the event that during the time between $\tau_j$ and $\tau_{j+1}$ the pair
$\{q_j , q_j + 1\}$ is added to $P$. The pair $\{q_j , q_j + 1\}$ is violating color $S(q_j)$. At time $\tau_j$, even conditioned
on all the past $P_1, \ldots , P_{\tau_j}$, in at most two steps (and before $\tau_{j+1}$) we can add the pair $\{q_j , q_j + 1\}$
to $P$. For every $j$, therefore,
\[
\Pr[E_j |P_1,\ldots , P_{\tau_j}] \geq (1/3)(1/3) = 1/9.
\]
Next, let $N' = \lceil qN/960\rceil$. We want to show that with high probability,  for at least $N'$ many $j$, the event $E_j$ occurs. There are $\binom{\lfloor qN/16\rfloor}{\lceil qN/960 \rceil}$ many ways of choosing a set of indices $j$ of size $N-N'$. Subsequently,
\begin{align*}
\Pr[\text{there is $j_1,\ldots , j_{N'}$ so that $E_{j_1}\cap \cdots \cap E_{j_{N'}}$}] &\geq  1 - \binom{\lfloor qN/16\rfloor}{\lceil qN/960 \rceil}\cdot \left(\frac{8}{9}\right)^{N-N'}\\
&\geq 1 - \left(\frac{960e}{16}\right)^{N'}\cdot \left(\frac{8}{9}\right)^{60N'}\\
&\geq 1 - c^{N'}
\end{align*}
where $0< c < 1$ is a universal constant. Finally, we argue that if there do exist $j_1,\ldots , j_{N'}$ for which the events $E_{j_1},\ldots, E_{j_{N'}}$ occur, then $|\calG(P)|\geq q/1000$. To see this, note that the size of every $Q_k$ is $N$. So, every color $k$ in $B$ can contribute at most $N$ elements to
$j_1, \ldots , j_{N'}$. If $|\calG(P)|<q/1000$, then at most these many colors can contribute more than $r^{2\eps}$ (and up to $N$ elements) - combined, at most $qN/1000$ elements. However, there are at least $q/2 - q/1000$ colors which can contribute only up to $r^{2\eps}$ elements. Again combined, this is not sufficient to cover the $N'$ elements overall (for large enough $d$ and a small enough constant $c_2$ that depends only on $\epsilon$), which is a contradiction. Hence,
\[
\Pr[|\calG(P)| \geq q/1000] \geq \Pr[\text{there is $j_1,\ldots , j_{N'}$ so that $E_{j_1}\cap \cdots \cap E_{j_N'}$}].
\]
and the proof follows in this case as $c^{N'}\ll r^{-\Omega(q)}$.

\paragraph{Case 2: Many colors with few jumps.}
The intuition is that many violations will come
from pairs of the form $\{L_t - 1, R_t + 1\}$ in the construction of the pairing.
Assume that for at least $q/2$ colors $k$,
$|J_k| \leq r^{4\eps}$.
Denote again by $B \subseteq [q]$ the set of $k$’s that satisfy the above inequality. We say that a color $k$ is
noticeable in the arc $A$ if
\[
r^{1 - 6\eps} \leq |S_{k}\cap A| \leq 
|A| - r^{1-6\eps}.
\]

\begin{claim}\label{clm:noticeable}
There are $q' \geq q/2 - 1$ disjoint arcs $A_1,\ldots, A_{q'}$ so that for every $j \in [q'
]$,
\begin{enumerate}
    \item $|A_j | = m = \lfloor r^{1-5\eps}\rfloor$ and,
\item there is a color $k_j$ in $B$ that is noticeable in $A_j$.
\end{enumerate}
Moreover, the colors $k_1,\ldots, k_{q'}$ can be chosen to be pairwise distinct.
\end{claim}

\begin{proof}
For each color $k$ in $B$, there are at least $r$ vertices of color $k$ in the $d$-cycle and at most
$r^{4\eps}$ jumps in the color $k$. Therefore, there is at least one $k$-monochromatic arc of size at least
$r^{1-4\eps}$. Hence, on the $d$-cycle, there are such monochromatic arcs $I_{k_1}
,\ldots, I_{k_{|B|}}$
for the colors
$k_1,\ldots, k_{|B|}$
in $B$, in this order $(1 < 2 <\cdots< D)$.

Consider an arc $A$ of size $m$ included in $I_{k_1}$. Thus $|S_{k_1} \cap A| = m$. If we “slide” the arc $A$ until
it is included in $I_{k_2}$, then $|S_{k_1} \cap A| = 0$. By continuity, there is an intermediate position for the arc
$A$ such that $r^{1-6\eps} \leq |S_{k_1} \cap A| \leq m - r^{1-6\eps}$. This provides the first arc $A_1$ of the claim.

Sliding an arc inside $I_{k_2}$
to inside $I_{k_3}$
shows that there exists an arc $A_2$ such that $r^{1-6\eps}\leq
|S_{k_2} \cap A_2| \leq m - r^{1-6\eps}$. The arcs $A_1$ and $A_2$ are disjoint: The distance of the largest element of $A_1$
and the smallest element of $I_{k_2}$
is at most $m$. The distance of the smallest element of $A_2$ and the
largest element of $I_{k_2}$
is at most $m$. The size of $I_{k_2}$
is larger than $2m$.
Proceed in this way to define $A_3,\ldots, A_{|B|-1}$.
\end{proof}

Use Claim \ref{clm:noticeable} to divide the construction of the (random) pairing into \emph{epochs}. Denote by $A^{(0)}$
the family of arcs given by the claim. Let $\tau_1$ be the first time $t$ that the arc $[L_t
, R_t
]$ hits one of
the arcs in $A^{(0)}$. Denote by $A_1$ that arc that is hit at time $\tau_1$ (break ties arbitrarily). Denote by
$k_1$ the color that is noticeable in $A_1$. Let $\sigma_1$ be the first time $t$ so that $A_1$ is contained in $[L_t
, R_t
]$.
Let $A^{(1)}$ be the subset of $A^{(0)}$ of arcs that have an empty intersection with $[L_{\sigma_1}
, R_{\sigma_1}
]$. Similarly,
let $\tau_2$ be the first time $t$ after $ \sigma_1$ that the arc $[L_t
, R_t
]$ hits one of the arcs in $A^{(1)}$. If there are no
arc in $A^{(1)}$, define $\tau_2 = \infty$. Denote by $A_2$ that arc that is hit at time $\tau_2$. Denote by $k_2$ the color that
is noticeable in $A_2$. 
Let $\sigma_2$ be the first time $t$ so that $A_2$ is contained in $[L_t, R_t]$. 
Let $A^{(2)}$ be the subset of $A^{(1)}$ of arcs that have an empty intersection with $[L_{\sigma_2}, R_{\sigma_2}]$. 
Define $\tau_j , \sigma_j , A_j , k_j, A^{(j)}$ for $j > 2$ analogously.
For every $j \geq 1$, denote by $E_j$ the event that during the time between $\tau_j$ and $\tau_{j+1}$ the number of
pairs added that violate color $k_j$'s at most $r^{2\eps}$. (If $E_j$ does not hold, then $|V_{k_j}
(P)| \geq r^{2\eps}$ and $k_j\in \calG(P)$.) 
The main part of the proof is summarized in the following proposition, whose proof is deferred to
Section \ref{subsec:chessboard}.

\begin{lem}[Chessboard Lemma]\label{lem:chessboard}
There is an absolute constant $0<\eps'\leq 1/100$ such that for every $j \geq 1$, and any choice of pairs $P_1,\ldots,P_{\tau_j}$,
\[\Pr[E_j|P_1,\ldots, P_{\tau_j}, |A^{(j-1)}| \geq 3]\leq r^{-\eps'}.\]
\end{lem}

Given this lemma, let us finish the proof of Lemma \ref{lem:many-viol}. Define $q'' = \lfloor q'/10 \rfloor$ and let $T$ denote the event that the number of $j$'s for which $|A^{(j)}|\geq 3$ is at least $q''$. First, we argue that $T$ occurs with high probability. 

For any $j\geq 1$, consider the evolution of the arc $[L_t,R_t]$ between the time steps $\tau_j$ (when it first hits arc $A_j$) and $\sigma_j$ (when it completely engulfs it). During this epoch, let us call the evolution of $[L_t,R_t]$ \emph{in} the `direction' of $A_j$ as \emph{good} (labelled `$G$') and \emph{away} from the direction of $A_j$ as \emph{bad} (`$B$'). To this end, for any time step in this epoch, we can \emph{code} the three possible choices for the evolution of $[L_t,R_t]$ as $GG$ (when the arc is grown \emph{in} the direction of $A_j$), $GB$ (when it is grown equally on either side), or $BB$ (when it is grown \emph{away} from the direction of $A_j$). Consequently, the evolution of $[L_t,R_t]$ during this epoch can be realized as a sequence consisting of the symbols $G$ and $B$.

Consider the sequence $s$ of $G$'s and $B$'s obtained by concatenating the sequences corresponding to all the epochs (ignoring the choices made at time steps that do \emph{not} lie in such epochs, i.e., between $\tau_j$ and $\sigma_j$ for some $j$ - as there is no corresponding notion of a `good' direction outside such epochs). The intuition is that if $|A^{(q'')}| < 3$ (i.e., if $T$ does not occur), then there must be an extremely large number of $B$'s compared to $G$'s (i.e., the arc $[L_t,R_t]$ evolves disproportionately in the \emph{bad} direction) in the concatenated string $s$, which should occur only with a vanishingly small probability.

Consider the sub-string $s'$ of $s$ that corresponds to the choices made only for the nodes in $A^{(0)}\setminus A^{(q'')}$. Note that there are precisely $mq''$ many $G$'s in $s'$. Suppose $|A^{(q'')}| = 2$ for concreteness (the cases $|A^{(q'')}| = 1$ and $|A^{(q'')}| = 0$ are similar). This implies that there are $m(q' - 2 - q'')$ many $B$'s in $s'$. Since only up to $mq''$ many of these $B$'s may appear as a result of the evolution making a choice of the form $GB$, it follows that the evolution of $[L_t,R_t]$ must make a choice of the form $BB$ at least $m(q' - 2 - 2q'')/2$ times out of a possible $m(q' - 2)/2$, in order to cover the elements of $A^{(0)}\setminus A^{(q'')}$. Denote $q_1 \coloneqq (q' - 2)/2$. By the union bound, this probability is at most 
\[
\Pr[|A^{(q'')}| = 2] \leq \binom{mq_1}{mq''}\cdot \left(\frac{1}{3}\right)^{m(q_1 - q'')} < c_2^{mq''}
\]
for some universal constant $0<c_2<1$. Similarly, we have bounds for both $\Pr[|A^{(q'')}| = 1]$ and $\Pr[|A^{(q'')}| = 0]$ and it follows that $\Pr[T] \geq 1 - c^{mq''}$ for some universal constant $0<c<1$. 

\begin{remark}
The argument above for showing that $T$ occurs with high probability differs considerably from \cite{DvirMPY12}, where the corresponding event is sketched to occur with probability only at least $1-dc^{m^{1/3}}$, which is not strong enough for our purposes.  
\end{remark}

Next, note that

\[
 \Pr[|\calG(P)| < q/1000] \leq \Pr[|\calG(P)|< q/1000 \cap T] + \Pr[\neg T]\leq \Pr[|\calG(P)|< q/1000 | T] + \Pr[\neg T].
\]
If $|\calG(P)| < q/1000$, then at least $q/2 - q/1000$ colors in $B$ have at most $r^{2\eps}$ many violations. Since $q'' = \lfloor q'/10\rfloor < q/2 - q/1000$, in particular, there must exist at least $q''/2$ colors within the \emph{first} $q''$ colors (here we are using the ordering of colors as provided by Claim \ref{clm:noticeable}) for which there are at most $r^{2\eps}$ many violations. We then obtain the following by conditioning on $T$, using the union bound.
\[
 \Pr[|\calG(P)|< q/1000 | T] \leq 2^{q''} \max_{H = \{j_1<\cdots< j_{q''/2}\} \subset [q'']} \Pr[E_{j_1},\ldots, E_{j_{q''/2}}||A^{(q'')}|\geq 3]
\]
For a fixed choice of $H$, by the chain rule and Lemma \ref{lem:chessboard}, we have
\begin{align*}
\Pr[E_{j_1}\cap \cdots \cap E_{j_{q''/2}}||A^{(q'')}|\geq 3] &= \Pr[E_{j_1}|T]\cdot \Pr[E_{j_2}|E_{j_1}\cap T]\cdot \cdots\cdot  \Pr[E_{j_{q''/2}}|E_{j_{q''/2 - 1}}\cap \cdots \cap E_{j_1} \cap T]\\ &\leq r^{-\eps' q''/2}\leq r^{-\eps' q'/20} \leq r^{-q/400}.
\end{align*}

Overall, setting $\beta$ appropriately, we conclude that 
\[
\Pr[|\calG(P)|< \alpha q] \leq r^{-\beta q}.
\]
\end{proof}

\subsection{Proof of the Chessboard Lemma}\label{subsec:chessboard}

To prove Lemma \ref{lem:chessboard}, we use a different point of view of the random process. We begin by describing this different view, and later describe its formal connection to the distribution on pairings. This subsection is adapted from Section 5 of \cite{DvirMPY12} and closely follows their argument, though with numerous parameter changes to suit our demands.

The view uses two definitions. One is a standard definition of a two-dimensional random
walk, and the other is a definition of a “chessboard” configuration in the plane. The proof of the
proposition will follow by analyzing the behavior of the random walk on the “chessboard”.
Let $d$ be as above and $m$ be as defined in Lemma \ref{clm:noticeable}. The random walk $W$ on $\N^2$
is defined as follows. It
starts at the origin, $W_0 = (0, 0)$. At every step it move to one of three nodes, independently of
previous choices,

\[
W_{t+1} = \begin{cases}
     W_t + (0,2) &\text{with probability $1/3$}\\
     W_t + (1,1) &\text{with probability $1/3$}\\
     W_t + (2,0) &\text{with probability $1/3$}
\end{cases}
\]

At time $t$, the $L_1$-distance of $W_t$ from the origin is thus $2t$.

The “chessboard” is defined as follows. Let $\alpha_1 : [m] \rightarrow \{0, 1\}$ and $\alpha_2 : [2m] \rightarrow \{0, 1\}$ be two
Boolean functions. The functions $\alpha_1, \alpha_2$ induce a “chessboard” structure on the board $[m] \times [2m]$.
A position in the board $\xi = (\xi_1, \xi_2)$ is colored either white or black. It is colored black if $\alpha_1(\xi_1) \neq
\alpha_2(\xi_2)$ and white if $\alpha_1(\xi_1) = \alpha_2(\xi_2)$. We say that the “chessboard” is well-behaved if

\begin{enumerate}
    \item $\alpha_1$ is far from constant:
    \[
r^{1-6\eps} \leq
|\{\xi_1 \in [m] : \alpha_1(\xi_1) = 1\}|  \leq m - r^{1-6\eps}.
    \]
    \item $\alpha_1$ does not contain many jumps:
    \[
|\{\xi_1 \in [m - 1] : \alpha_1(\xi_1) \neq \alpha_1(\xi_1 + 1)\}|\leq r^{4\eps}
    \]
    \item $\alpha_2$ does not contain many jumps:
   \[
|\{\xi_2 \in [2m - 1] : \alpha_2(\xi_2) \neq \alpha_2(\xi_2 + 1)\}|\leq r^{4\eps}
    \]
\end{enumerate}

Consider a random walk $W$ on top of the “chessboard” and stop it when reaching the boundary
of the board (i.e., when it tries to make a step outside the board $[m] \times [2m]$). We define a good
step to be a step of the form $(1, 1)$ that lands in a black block. We will later relate good steps to
violating edges. Our goal is, therefore, to show that a typical $W$ makes many good steps.

\begin{lem}\label{lem:good-steps}
    Assume the chessboard is well-behaved. There is a constant $0<\eps'\leq 1/100$ such that the probability that $W$ makes less than $r^{4\eps}$
good steps is at most $r^{-\eps'}$.
\end{lem}

We use this lemma to show Lemma \ref{lem:chessboard}.

\begin{proof}[Proof of Lemma \ref{lem:chessboard} given Lemma \ref{lem:good-steps}]
    Recall that $A_j$ is an arc of size $|A_j| = m = \lfloor r^{1-\delta - \delta'}\rfloor$
 so that there is a color
$k_j$ satisfying
\begin{equation}\label{eq:noticeable}
r^{1-6\eps} \leq |S_{k}\cap A| \leq 
|A| - r^{1-6\eps}.
\end{equation}

Furthermore, condition on $P_1, \ldots, P_{\tau_j}$
, $|A^{(j-1)}| \geq 3$. Assume without loss of generality that $R_{\tau_j}$
is in $A_j$ (when $L_{\tau_j}$
is
in $A_j$, the analysis is similar). The distance of $R_{\tau_j}$
from the smallest element of $A_j$ is at most one
(the length of “one step to the right” is between zero and two). We now grow the random interval
until $\sigma_j$, i.e., as long as $R_t$ stays in $A_j$. At the same time, $L_t$ performs a movement to the left.
Since $|A^{(j-1)}| \geq 3$, there are at least $2m$ steps for $L_t$ to take to the left before hitting $A_j$.
There is a one-to-one correspondence between pairings $P$ and random walks $W$ using the correspondence
\[
P_{t+1} = \{L_t - 2, L_t - 1\} \longleftrightarrow W_{t+1} = W_t + (0, 2),
\]
\[
P_{t+1} = \{L_t - 1, R_t + 1\} \longleftrightarrow W_{t+1} = W_t + (1, 1), \]
\[
P_{t+1} = \{R_t + 1, R_t + 2\} \longleftrightarrow W_{t+1} = W_t + (2, 0).
\]

Define the function $\alpha_1$ to be $1$ at positions of $A_j$ with color $k_j$, and $0$ at the other positions. Set
the function $\alpha_2$ as to describe the color $k_j$ from $L_{\tau_j}$
leftward. The “chessboard” is well-behaved by
(\ref{eq:noticeable}) and since $k_j$ is in the set $B$ defined in case 2 of the proof of Lemma \ref{lem:many-viol} (so there are not many
jumps for the color $k_j$).
Finally, if $W$ makes a good step, then the corresponding pair added to $P$ violated color $k_j$ .
So, if $E_j$ holds for $P$, then the corresponding $W$ makes less than $r^{4\eps}$ good steps. Formally, by
Lemma \ref{lem:many-viol},

\[
\Pr[E_j|P_1,\ldots, P_{\tau_j}, |A^{(j-1)}| \geq 3]\leq \Pr[\text{$W$ makes less than $r^{4\eps}$ good steps}] \leq r^{-\eps'}.
\]
\end{proof}

\begin{proof}[Proof of Lemma \ref{lem:good-steps}]
    Define three events $E_R, E_C, E_D$, all of which happen with small probability, so that every $W$ that
is not in their union makes many good steps.

Call a subset of the board of the form $I \times [2m]$ or $[m] \times I$, where $I$ is a sub-interval, a \emph{region}.
The width of a region is the size of $I$. Let $R$ be the set of regions of width at least $r^{8\eps}$. The size
of $R$ is at most $2m^2$. For a region $t$ in $R$, denote by $E_t$ the event that the number of steps of the
form $(1, 1)$ that $W$ makes in $t$ is less than $r^{4\eps}$ \emph{given} that it makes at least $r^{6\eps}$ steps in $r$. Denote

\[
E_R = \bigcup_{t\in R} E_t
\]

To estimate the probability of $E_t$, note that we can simply apply the Chernoff bound to a sum of $r^{6\eps}$ Bernoulli random variables with $p = 1/3$. By the union bound, we conclude that there is a universal constant $0<c<1$ such that

\[
\Pr[E_R] \leq c^{r^{6\eps}}.
\]

Denote by $H$ the set of all points in the board with $L_1$-norm at least $m^{5/8}$. At time $T$ the
random walk $W$ is distributed along all points in $\N^2$ of $L_1$-norm exactly $T$. The distribution of
$W$ on this set is the same as that of a random walk on $\Z$ that is started at $0$, and moves at every
step to the right with probability $1/3$, stays in place with probability $1/3$ and moves to the left with probability $1/3$. The probability
that such a random walk on $\Z$ is at a specific point in $\Z$ at time $T$ is at most $O(T^{-1/2})$. Hence, for
every point $h$ in $H$,

\[
\Pr[W \text{ hits } h]\leq O(m^{-5/16}).
\]

Call a point $c = (\xi_1, \xi_2)$ in the board a corner if both $(\xi_1, \xi_2)$ and $(\xi_1 + 1, \xi_2 + 1)$ are of the same
color $\kappa \in \{\mathrm{black}, \mathrm{white}\}$, but $(\xi_1 + 1, \xi_2)$ and $(\xi_1, \xi_2 + 1)$ are not of color $\kappa$. For a corner $c$, denote
by $\Delta(c)$ the $r^{8\eps}$-neighborhood of $c$ in $L_1$-metric. Denote by $\Delta$ the union over all $\Delta(c)$, for corners
$c$ in $H$. Denote by $E_C$ the event that $W$ hits any point in $\Delta$. Since the board is well-behaved, the
number of jumps in each of $\alpha_1, \alpha_2$ is at most $r^{4\eps}$. Therefore, the number of corners is at most
$r^{8\eps}$. By the union bound,
\[
\Pr[E_C] \leq O(r^{8\eps}r^{16\eps}m^{-5/16}) \leq r^{-\eps'},
\]
where in the last step, we used $m \geq r^{1 - 5\eps}$. Note that plugging in, say, $\eps = \eps' = 1/100$ indeed makes the inequality true. Next, let $m' = \lceil m^{5/8}\rceil$. Define three (vertical) lines: $D_1$ is the line $\{m'\} \times [2m]$, $D_2$ is the line
$\{2m'\} \times [2m]$ and $D_3$ is the line $\{m - m'\} \times [2m]$. Denote by $E_D$ the event that $W$ does not cross
the line $D_3$ before stopped (i.e., hitting the boundary of the board). Chernoff’s bound implies that there is a universal constant $0<c<1$ for which
\[
P[E_D] \leq c^m.
\]

To conclude the proof by the union bound, it suffices to show that for every $W$ not in $E_R \cup
E_C \cup E_D$, the walk $W$ makes at least $r^{4\eps}$ good steps. Fix such a walk $W$. Since $W \notin E_D$, we
know that $W$ crosses the line $D_2$.

We consider several cases. Define a \emph{block} to be a maximal monochromatic rectangle in the board.
The board is thus partitioned into black blocks and white blocks - which is what led \cite{DvirMPY12} to calling it a  “chessboard.”
We now think of the board $[m] \times [2m]$ as drawn in the plane with $(1, 1)$ at the bottom-left corner and
$(m, 2m)$ at the upper-right corner.

\textbf{Case 1:}  The walk $W$ does not hit any white block after crossing $D_1$ and before crossing $D_2$.
In this case, all steps taken in the \emph{region} whose left border is $D_1$ and right border is $D_2$ are in a black
area. The number of such steps is at least $m^{5/8}/2\gg r^{6\eps}$. Since $W \notin E_R$, the claim holds.

\textbf{Case 2:}  The walk W hits a white block after crossing $D_1$ and before crossing $D_2$. Let us label the blocks as follows: we associate every block with a pair $\langle \eta_1, \eta_2\rangle$ where $\eta_1$ is between $1$ and the number of jumps in $\alpha_1$ and $\eta_2$ is between $1$ and the number of jumps
in $\alpha_2$. So, the label of the ``bottom-left'' is $\langle 1,1\rangle$, the label of the block ``above'' it is $\langle 1,2\rangle$ and the label of the block ``to its right'' is $\langle 2,1\rangle$, etc.  There are
two sub-cases to consider: 

\textbf{Sub-case 1:} At some point after crossing $D_1$ and before crossing $D_3$, there are two white blocks
of the form $\langle \eta_1, \eta_2\rangle$, $\langle\eta_1 + 1, \eta_2 + 1\rangle$ so that $W$ intersects
both blocks. Let $c$ be the corner between these two blocks (which must exist by definition). Since
$W \notin E_C$, we know that $W$ does not visit $\Delta(c)$. Therefore, $W$ must walk in a black area around
$\Delta(c)$. Every path surrounding $\Delta(c)$ has length at least $r^{8\eps}$. Since $W \notin E_R$, the claim holds.

\textbf{Sub-case 2:} At all times after crossing $D_1$ and before crossing $D_3$, the walk never moves from a
white block $\langle \eta_1, \eta_2 \rangle$ to one of the two white blocks $\langle\eta_1 + 1, \eta_2 + 1\rangle$, $\langle \eta_1 - 1, \eta_2 - 1\rangle$. Since $W \notin E_D$,
this is indeed the last case. The width of a combinatorial rectangle in the board is the size of its
“bottom side” (i.e., the corresponding subset of $[m]$). Let $\eta$ be the first white block $W$ hits after
crossing $D_1$. Let $\Sigma$ be the family of black blocks that are to the right but on the same height as $\eta$.
Define $Z$ as the maximal width of a rectangle of the form $\sigma \cap [0, m - m_0 - 1] \times [2m]$ over all $\sigma \in \Sigma$.
Since the board is well-behaved, it follows (from the first condition) that the
total width of the black area on the same height as $\eta$ is at least $r^{1-6\eps}$. Also, since we are in case 2, the left border of $\eta$ is to the left of $D_2$. Therefore, the
total width of the black area to the right of the left border of $\eta$ and to the
left of $D_3$, on the same height as $\eta$ is at least $r^{1-6\eps}- 3m'$. Therefore, since the number of jumps is at most $r^{4\eps}$,
\[
Z\geq (r^{1-6\eps}- 3m')/r^{4\eps} \gg r^{8\eps}.
\]

Since we are in this sub-case, the walk $W$ must “go through” every black block it hits: it can go from
bottom side to upper side or from left side to right side (but not from left side to upper side or
from bottom side to right side). Consider the behaviour of $W$ after it hits $\eta$: starting from a white block, because $W \notin E_D$, it is guaranteed to cross $D_3$. Therefore, the color of the block that $W$ ``exits'' from from each \emph{column} must keep alternating between white and black. For each black block in $\Sigma$, therefore, there exists
a black block in the same column that $W$ crosses horizontally. Focusing on one such black block
of width $Z$, since $W \notin E_R$, the claim holds.
\end{proof}

\end{document}